\documentclass[aps,prb,twocolumn,superscriptaddress]{revtex4-1}
\usepackage{amsmath,amsthm,amssymb}
\usepackage[dvipdfmx]{graphicx}
\usepackage{amsmath,bm}
\usepackage{color,ulem}
\usepackage{ifthen}

\newcount \commentout
\newcommand{\1}{\mbox{1}\hspace{-0.25em}\mbox{l}}

\newlength{\figwidth}
\newlength{\figlarge}
\begin{document}
%%%%%%%%%%%%%%%%%%%%%%%%%%%%%%%%%%%%%%%%%%%%%%%%%%%%%%%%%%%%%%%%%%%%%%%
\title{
Fate of Majorana modes in $\mathrm{CeCoIn}_5$/$\mathrm{YbCoIn}_5$ superlattices\\
%Topology of a superconductor in superlattice of $\mathrm{CeCoIn}_5$/$\mathrm{YbCoIn}_5$\\ 
-a test bed for reduction of topological classification-
}
%%%%%%%%%%%%%%%%%%%%%%%%%%%%%%%%%%%%%%%%%%%%%%%%%%%%%%%%%%%%%%%%%%%%%%%
\author{Tsuneya Yoshida}
\affiliation{Department of Physics, Kyoto University, Kyoto 606-8502, Japan}
\author{Akito Daido}
\affiliation{Department of Physics, Kyoto University, Kyoto 606-8502, Japan}
\author{Youichi Yanase}
\affiliation{Department of Physics, Kyoto University, Kyoto 606-8502, Japan}
\author{Norio Kawakami}
\affiliation{Department of Physics, Kyoto University, Kyoto 606-8502, Japan}
%%%%%%%%%%%%%%%%%%%%
%%%%%%%%%%%%%%%%%%%%%%%%%%%%%%%%%%%%%%%%%%%%%%%%%%%%%%%%%%%%%%%%%%%%%%%
\date{\today}
%%%%%%%%%%%%%%%%%%%%%%%%%%%%%%%%%%%%%%%%%%%%%%%%%%%%%%%%%%%%%%%%%%%%%%%
\begin{abstract}
%In this paper, we propose  $\mathrm{CeCoIn_5}/\mathrm{YbCoIn_5}$ superlattice systems as a test bed for the reduction of topological classification in free fermions. 
%\magenta{In the superlattice systems, the number of $\mathrm{CeCoIn_5}$ layers can be experimentally tuned.}
%We find that the system with quad-layer of $\mathrm{CeCoIn_5}$ shows a topological crystalline superconducting phase with the mirror Chern number eight at the non-interacting level.
%Furthermore, \magenta{our bosonization analysis} demonstrates that in the presence of two-body interactions, gapless edge modes are no longer protected by the symmetry in the system with quad-layer, but are protected in the systems with bi- or tri-layer. This clearly exemplifies the reduction of topological classification from $\mathbb{Z}$ to $\mathbb{Z}_8$.
In this paper, we propose $\mathrm{CeCoIn_5}/\mathrm{YbCoIn_5}$ superlattice systems as a test bed for the reduction of topological classification in free fermions. 
We find that the system with quad-layer of $\mathrm{CeCoIn_5}$ shows a topological crystalline superconducting phase with the mirror Chern number eight at the non-interacting level. 
Furthermore, we demonstrate that in the presence of two-body interactions, gapless edge modes are no longer protected by the symmetry in the system with quad-layer, but are protected in the systems with bi- or tri-layer. This clearly exemplifies the reduction of topological classification from $\mathbb{Z}\oplus\mathbb{Z}$ to $\mathbb{Z}\oplus\mathbb{Z}_8$.
\end{abstract}
%%%%%%%%%%%%%%%%%%%%%%%%%%%%%%%%%%%%%%%%%%%%%%%%%%%%%%%%%%%%%%%%%%%%%%%
\pacs{
***
}
%%%%%%%%%%%%%%%%%%%%%%%%%%%%%%%%%%%%%%%%%%%%%%%%%%%%%%%%%%%%%%%%%%%%%%%
%%%%%%%%%%%%%%%%%%%%%%%%%%%%%%%%%%%%%%%%%%%%%%%%%%%%%%%%%%%%%%%%%%%%%%%
\maketitle
%%%%%%%%%%%%%%%%%%%%%%%%%%%%%%%%%%%%%%%%%%%%%%%%%%%%%%%%%%%%%%%%%%%%%%%
%%%%%%%%%%%%%%%%%%%%%%%%%
\textit{-Introduction-}
%%%%%%%%%%%%%%%%%%%%%%%%%
After the discovery of the topological insulators (TIs) and topological superconductors (TSCs), topological properties of quantum phases have been extensively studied\cite{TI_review_Hasan10,TI_review_Qi10}. 
In TIs/TSCs, nontrivial topology of the wave function in the bulk predicts gapless excitations at boundary/surface of the systems, which are sources of novel transport properties. 
These systems are free fermion systems (i.e., they are described by a quadratic Hamiltonian), and the topology of TIs/TSCs is protected by symmetries. 
Examining how many topological phases exist under a given local symmetry (i.e., classification of TIs and TSCs) is an important issue and gives useful information\cite{Schnyder_classification_free_2008,Kitaev_classification_free_2009,Ryu_classification_free_2010}. 
On the experimental side, realization of TIs/TSCs has been a significant issue, and various numbers of TIs/TSCs have been indeed realized; a two-dimensional TI was first confirmed for a quantum well of $\mathrm{HgTe}$/$\mathrm{CdTe}$\cite{HgTe_Bernevig06,exp_2D-QW_MKonig_2007}, and three-dimensional TIs were reported for $\mathrm{Bi_2Se_3}$ \textit{etc.}\cite{exp_3D-bismuth_YXia_2008,exp_3D-bismuth_YLChen,exp_3D_bismuth_Se_Zhang}. Also, a TSC was proposed for $\mathrm{Cu_xBi_2Se_3}$\cite{exp-3D_TSC_Sasaki11}.

Understanding the effects of electron correlations, which are generally neglected in the treatment of TIs/TSCs, is one of the current important issues in this field. Theoretical proposals of TIs in strongly correlated compounds have further stimulated this issue\cite{NaIrO_Nagaosa09,Heusler_Chadov10,Heusler_Lin10,skutterudites_Yan12,Takimoto_2011,SmB6_Lu2013,Kargarian_Fiete_TCIinOxide2013,Hsieh_TCIinOxides2014,Weng_Dai_TCIinYbB12_2014}.
Recent extensive studies have discovered  the reduction of topological classification for free fermion systems. In particular,
Fidkowski and Kitaev have found that for a one-dimensional TSC of class BDI, the topological classification for free fermions, $\mathbb{Z}$, collapses into $\mathbb{Z}_8$ in the presence of electron correlations\cite{Z_to_Zn_Fidkowski_10,Fidkowski_1Dclassificatin_11,Turner11}.
This means that gapless edge modes can be unstable against electron interactions.
The reduction of topological classification has further been extended to two- and three-dimensional systems by examining stability/instability of gapless edge modes\cite{Lu_CS_2011,Levin_CS_2012,YaoRyu_Z_to_Z8_2013,Ryu_Z_to_Z8_2013,Hsieh_CS_CPT_2014,Wang_Potter_Senthil2014,Wang_Senthil2014,You_Cenke2014,Isobe_Fu2015,Yoshida2015,Morimoto_2015,TMKI_Zhang2016,Yoshida_2017}.
In spite of such significant progress on the theoretical side, the reduction of topological classification has not been experimentally reported yet. 
Therefore, an experimental platform is indispensable for progress in this direction. 

In this paper, we address the following question. 
\textit{How can we realize an experimentally accessible test bed to observe the reduction of the classification?} 
In order to answer this question, we analyze topological properties of superconducting phases in a superlattice system composed of $\mathrm{CeCoIn_5}/\mathrm{YbCoIn_5}$ layers, which is the only known example of experimentally realizable two-dimensional heavy-fermion systems\cite{Mizukami_CeCoIn5YbCoIn5_11,Goh_superlattice12,Shimozawa_superlattice_PRL14,Shimozawa_superlattice_RPP2016}.
Topological properties of $s$-wave superconductors in the bi- and tri-layer have been studied at the non-interacting level\cite{superlattice_Yanase15}, which have been extended to the case of $d$-wave\cite{superlattice_Yanase16}. 
%\green{
%Topological properties of superconductors in the bi- and tri-layer have been studied at the non-interacting level\cite{superlattice_Yanase15,superlattice_Yanase16}. 
%}
In this paper, firstly, we demonstrate that a topological crystalline superconductor emerges in the system with quad-layer of $\mathrm{CeCoIn_5}$, following the spirits of these studies for multilayer superconductors. These topological crystalline superconductors form an Abelian group $\mathbb{Z}\oplus\mathbb{Z}$.
Secondly, our bosonization approach elucidates that the topological classification $\mathbb{Z}\oplus\mathbb{Z}$ for free fermions collapses into $\mathbb{Z}\oplus\mathbb{Z}_8$ in the presence of electron correlations.
These results suggest that the $\mathrm{CeCoIn_5}/\mathrm{YbCoIn_5}$ superlattice system provides an experimental test bed for the reduction of topological classification for free fermions. Besides that, we find that the number of $\mathrm{CeCoIn_5}$ layers is essential for the reduction; the reduction occurs in the system with quad-layer of $\mathrm{CeCoIn_5}$, although it does not in the systems with bi- or tri-layer of $\mathrm{CeCoIn_5}$.

\begin{figure}[!h]
\begin{center}
\includegraphics[width=60mm,clip]{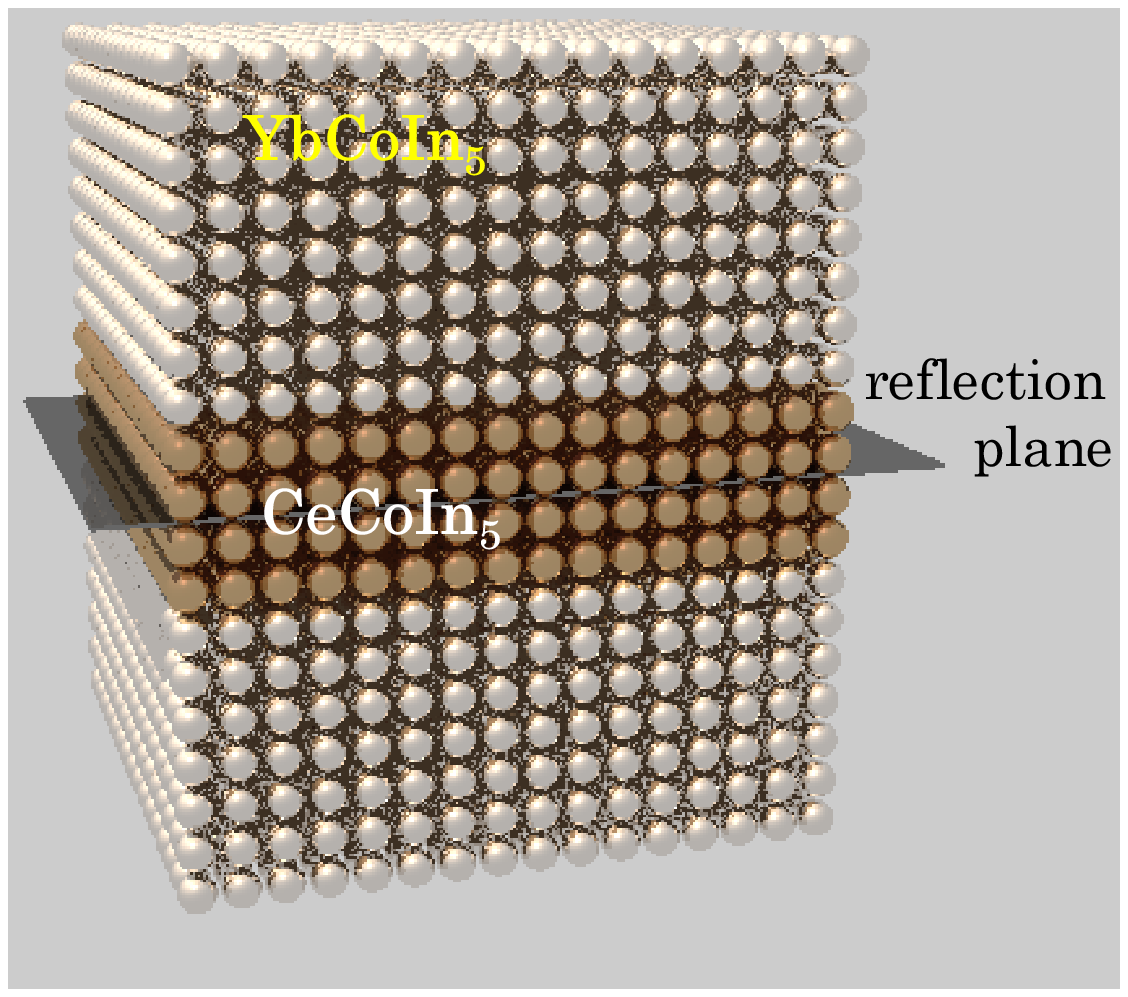}
\end{center}
\caption{(Color Online). 
Sketch of the superlattice. The quad-layer of $\mathrm{CeCoIn_5}$ (brown sphere) is sandwiched by $\mathrm{YbCoIn_5}$ (white sphere).
The reflection plane (black plane) is parallel with the two-dimensional sheet of $\mathrm{CeCoIn_5}$. This reflection plane locates between layers, which yields the Rashba spin-orbit coupling.
}
\label{fig:model}
\end{figure}

%%%%%%%%%%%%%%%%%%%%%%%%%
\textit{-Superlattice with quad-layer of $\mathrm{CeCoIn_5}$-}
%%%%%%%%%%%%%%%%%%%%%%%%%
We consider a system of the $\mathrm{CeCoIn_5}/\mathrm{YbCoIn_5}$ superlattice (Fig.~\ref{fig:model}).
Experimentally, thickness of $\mathrm{CeCoIn_5}$ and $\mathrm{YbCoIn_5}$ can be tuned at the atomic scale~\cite{Mizukami_CeCoIn5YbCoIn5_11,Goh_superlattice12,Shimozawa_superlattice_PRL14,Shimozawa_superlattice_RPP2016}. 
Besides that, the heavy fermions are confined in two-dimensional $\mathrm{CeCoIn_5}$ layers because the proximity effects are suppressed by large mismatch in the Fermi velocity between $\mathrm{CeCoIn_5}$ and $\mathrm{YbCoIn_5}$ \cite{superlattice_proximity_12,superlattice_Yanase_12}. 
Therefore, we focus on the subsystem of $\mathrm{CeCoIn_5}$. Here, in particular, we discuss the case of the quad-layer.

%%%%%%%%%%%%%%%%%%%%%%%%%
%%%%%%%%%%%%%%%%%%%%%%%%%

%%%%%%%%%%%%%%%%%%
\begin{subequations}
%%%%%%%%%%%%%%%%%%
\label{eq: setup}
\begin{eqnarray}
H &=& \sum_{\bm{k},m,\sigma,\sigma'}
c^\dagger_{\bm{k}m\sigma} 
[\hat{h}_m(\bm{k})]_{\sigma\sigma'}
c_{\bm{k}m\sigma'} 
\nonumber \\
&&
+\sum_{\bm{k},\langle mm'\rangle,\sigma}
t_\bot
c^\dagger_{\bm{k}m\sigma} 
c_{\bm{k}m'\sigma}+h.c.
\nonumber \\
&&
+
\sum_{\bm{k},\sigma,\sigma'}
\Delta_{m\sigma\sigma'}(\bm{k})
c^\dagger_{\bm{k}m\sigma}
c^\dagger_{-\bm{k}m\sigma'}
+h.c.,
\end{eqnarray}
%%%%%%%%%%%%%%%%%%
\end{subequations}
%%%%%%%%%%%%%%%%%%
where $c^\dagger_{\bm{k}m\sigma}$ is the creation operator of an electron at layer $m=1,2,3,4$ in a state with momentum $\bm{k}:=(k_x,k_y)$ and spin $\sigma=\uparrow,\downarrow$.
The first term denotes the normal part of the Hamiltonian for each layer, while the second term denotes the hopping between the neighboring layers.
The Hamiltonian locally breaks the reflection symmetry (i.e., the layer $m$ is mapped to the layer $5-m$.),
which leads to the spin-orbit interaction of Rashba type\cite{Goh_superlattice12,Shimozawa_superlattice_PRL14,Shimozawa_superlattice_RPP2016}. 
The last term represents the pairing potential of superconductors.
In the presence of the magnetic field, the matrix $\hat{h}_m(\bm{k})$ in the first term is written as 
$\hat{h}_m(\bm{k})= \xi(\bm{k})\sigma^0 +\alpha_m \bm{g}(\bm{k})\cdot \bm{\sigma}-\mu_{B}H\sigma^z$
with $\xi(\bm{k}):=-2t\left( \cos(k_x)+\cos(k_y) \right)-\mu$, where $t$ and $\mu$ denote the hopping strength and the chemical potential, respectively.
The second term represents the spin-orbit interaction, and $\bm{g}(\bm{k}):=(-\sin(k_y),\sin(k_x),0 )^T$.
Due to the symmetry breaking of local reflection, the singlet and triplet pairing states are mixed. Hence, the pairing potential is written as 
$\Delta_m(\bm{k})= i\left(\psi_m(\bm{k})-\bm{d}_m(\bm{k})\cdot \bm{\sigma} \right)\sigma^y$.
We note that the system is invariant under the reflection which maps electrons in the layer $m$ to those in the layer $5-m$.
Furthermore, the superconducting phase in layers of $\mathrm{CeCoIn_5}$ is supposed to be a pair-density-wave (PDW) phase because the superconductor in this system is 
(i) in the Pauli limit\cite{Onuki_CeCoIn5_Hc2_02}, 
(ii) quasi-two-dimensional, 
(iii) affected by strong spin orbit coupling\cite{Goh_superlattice12,Maruyama_superlattice_12}. 
Therefore, the parameter $\alpha$ and the pairing potential for each layer $\left( \alpha_m,\psi_m(\bm{k}),\bm{d}_m(\bm{k})\right)$ are assigned as
%%%%%%%%%%%%%%%%%%
\begin{subequations}
\begin{eqnarray}
\left( \alpha_1,\psi_1(\bm{k}),\bm{d}_1(\bm{k})\right) &=& \left( \alpha,\psi(\bm{k}),\bm{d}(\bm{k}) \right), \\
\left( \alpha_2,\psi_2(\bm{k}),\bm{d}_2(\bm{k})\right) &=& \left( \alpha',\psi'(\bm{k}),\bm{d}'(\bm{k}) \right), \\
\left( \alpha_3,\psi_3(\bm{k}),\bm{d}_3(\bm{k})\right) &=& \left( -\alpha',-\psi'(\bm{k}),\bm{d}'(\bm{k}) \right), \\
\left( \alpha_4,\psi_4(\bm{k}),\bm{d}_4(\bm{k})\right) &=& \left( -\alpha,-\psi(\bm{k}),\bm{d}(\bm{k}) \right).
\end{eqnarray}
\end{subequations}
%%%%%%%%%%%%%%%%%%
We note that in the bulk of $\mathrm{CeCoIn}_5$, $d_{x^2-y^2}$-wave is dominant\cite{angle_resolved_thermal_Matsuda_06}. 
Besides that, the most primitive triplet pairing is $p$-wave which mixes with the $d_{x^2-y^2}$-wave due to the Rashba interaction.
Based on these results and the analysis with group theory\cite{SigristUeda_91} (for details see Sec.~I of Ref.~\onlinecite{supp}), we conclude that the pairing potential is written as
$\psi(\bm{k}):=\Delta_{d}(\cos(k_x)-\cos(k_y))$ and 
$
\bm{d}(\bm{k}): =
c
\left(
\begin{array}{ccc}
\sin(k_y), & \sin(k_x), & 0
\end{array}
\right)^T
$, where $\Delta_{d}$ and $c$ are real numbers.
$\psi'(\bm{k})$ and $\bm{d}'(\bm{k})$ are defined in a similar way, just by replacing $(\Delta_d,c)$ with $(\Delta'_d,c')$.\\

By using the Nambu operator, the Hamiltonian is written as $H =\frac{1}{2}\sum_{\bm{k}}\Psi^\dagger_{\bm{k}} \mathcal{H}(\bm{k}) \Psi_{\bm{k}}$, where $\Psi_{\bm{k}}$ is 
the Nambu operator, $\Psi_{\bm{k}}:=\oplus_{\sigma}( c_{\bm{k}1\sigma},\cdots,c_{\bm{k}4\sigma}, c^\dagger_{-\bm{k}1\sigma},\cdots,c^\dagger_{-\bm{k}4\sigma} )^T$.
Since the system respects the reflection symmetry, the Bogoliubov-de Gennes (BdG) Hamiltonian $\mathcal{H}(\bm{k})$ can be block diagonalized in each eigenspace with an eigenvalue $\lambda i$ $(=+i,-i)$:
%%%%%%%%%%%%%%%%%%
\begin{eqnarray}
\mathcal{H}_{\lambda}(\bm{k})&=& 
\left(
\begin{array}{cc}
\hat{h}_\lambda(\bm{k}) &  \hat{\Delta} (\bm{k}) \\
\hat{\Delta}^\dagger(\bm{k}) & -\hat{h}^T_\lambda(-\bm{k})
\end{array}
\right),
\label{eq: BdG}
\end{eqnarray}
where $\hat{h}_\lambda(\bm{k})$ and $\hat{\Delta}(\bm{k})$ are four dimensional matrices, whose definition is given in Sec.~II of Ref.~\onlinecite{supp}.

\textit{-Topological class and topological invariants-}
Let us turn to the topological properties of the Hamiltonian. 
We note that the Hamiltonian matrix~(\ref{eq: BdG}) has a gapful spectrum due to the inter-layer hopping $t_\bot$.
The topological class of the Hamiltonian, $\mathcal{H}_{\lambda}(\bm{k})$, is class D according to the periodic table of TIs/TSCs \cite{Schnyder_classification_free_2008,Kitaev_classification_free_2009,Ryu_classification_free_2010,shiozaki_classification_2014}. This can be checked as follows.
The block diagonalized Hamiltonian $\mathcal{H}_{\lambda}(\bm{k})$ respects the reflection and particle-hole symmetry whose matrix representation is denoted by $\mathcal{R}$ and $\mathcal{U}_c\mathcal{K}$, respectively, where
 the operator $\mathcal{K}$ takes complex conjugation. For the explicit representation of these symmetry transformations, see Sec.~II of Ref.~\onlinecite{supp}.
Each operator satisfies the following relations: (i) $\mathcal{R}^2=-\1$; (ii) $\mathcal{R}$ and $\mathcal{C}$ anti-commute, $\{\mathcal{R},\mathcal{C}\}=0$.
Therefore, the topological class of the Hamiltonian matrix $\mathcal{H}_{\lambda}(\bm{k})$ is class D.

The topological properties of BdG Hamiltonian are characterized by the total Chern number $\nu_\mathrm{tot}$ and the mirror Chern number $\nu_\mathrm{M}$ of $\mathcal{H}(\bm{k})$.
These are linear combination of Chern numbers $\nu_{+}$ and $\nu_{-}$ for block diagonalized Hamiltonians, $\mathcal{H}_{+}(\bm{k})$ and $\mathcal{H}_{-}(\bm{k})$, whose topological class is D,
%%%%%%%%%%%%%%%%%%
\begin{eqnarray}
\nu_\mathrm{tot}= \nu_{+}+\nu_{-}, &\quad& \nu_\mathrm{M}= \frac{\nu_{+}-\nu_{-}}{2}.
\label{eq: nu_tot}
\end{eqnarray}
%%%%%%%%%%%%%%%%%%
(The definition of $\nu_{+}$ and $\nu_{-}$ are given in Sec.~II of Ref.~\onlinecite{supp}.) 
Namely, the topological crystalline superconducting phases form an Abelian group $\mathbb{Z}\oplus\mathbb{Z}$.
The total Chern number indicates a difference in the number of right- and left-movers of edge states. 
Thus, the system shows $\nu_\mathrm{M}$ helical edge modes for $\nu_\mathrm{tot}=0$.

\textit{-Topological properties of the BdG Hamiltonian-}
Now we characterize the topological properties of the system at the non-interacting level.
We set $t=1.0$, $t_{\bot}=0.1$, $\alpha=0.3$, $\alpha'=0.2$, $\Delta_d=0.05$, $\Delta'_d=0.05$, $c=-0.01$, and $c'=-0.0067$.
%\blue{
We note that essentially the same results are obtained in other cases of the parameter set.
%}

\begin{figure}[!h]
\begin{center}
\includegraphics[width=90mm,clip]{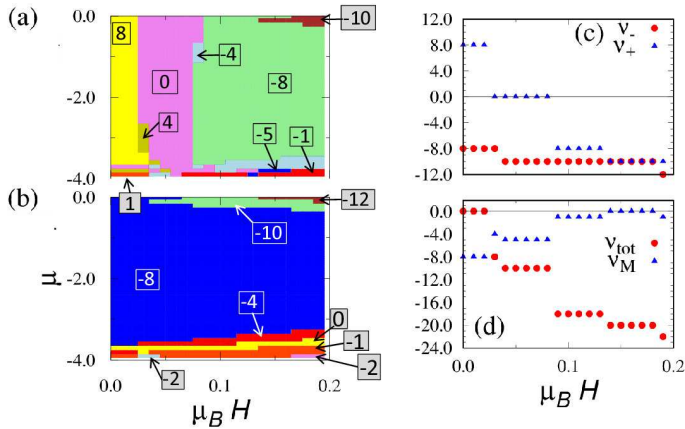}
\end{center}
\caption{(Color Online). 
(a) and (b) Chern number of $\mathcal{H}_+$ and $\mathcal{H}_-$. 
Numbers enclosed with squares denote the Chern number $\nu_\pm$.
We choose the following parameter set:
$t=1.0$, 
$t_{\bot}=0.1$, 
$\alpha=0.3$ $\alpha'=0.2$,
$\Delta_d=\Delta'_d=0.05$,
$c=-0.01$, and $c'=-0.0067$.
(c) Chern number for each sector as a function of the magnetic field for $\mu=-0.1$.
(d) Total Chern number and the mirror Chern number as a function of the magnetic field for $\mu=-0.1$. 
We choose the same parameter set as panel (a) and (b).
}
\label{fig: ch}
\end{figure}

In Figs.~\ref{fig: ch}(a) and (b), the Chern numbers for each subsector are plotted as functions of the chemical potential and the magnetic field.
(For calculation of the Chern numbers we employ the method proposed by Fukui \textit{et al.}\cite{Fukui_Hatsugai_05}.)
In the absence of the magnetic field the system is time-reversal invariant. Correspondingly, the total Chern number is zero, while the mirror Chern number is eight [see also Figs.~\ref{fig: ch}(c) and (d)].
Under the weak magnetic field, these two topological invariants have the same value as in the case of $\mu_B H=0$. 
Further increasing the magnetic field changes the sign of $\nu_+$, and the Chern number becomes $-16$.

%%%%%%%%%%%%%%%%%%%%%%%%%
\begin{figure}[!h]
\begin{center}
\includegraphics[width=80mm,clip]{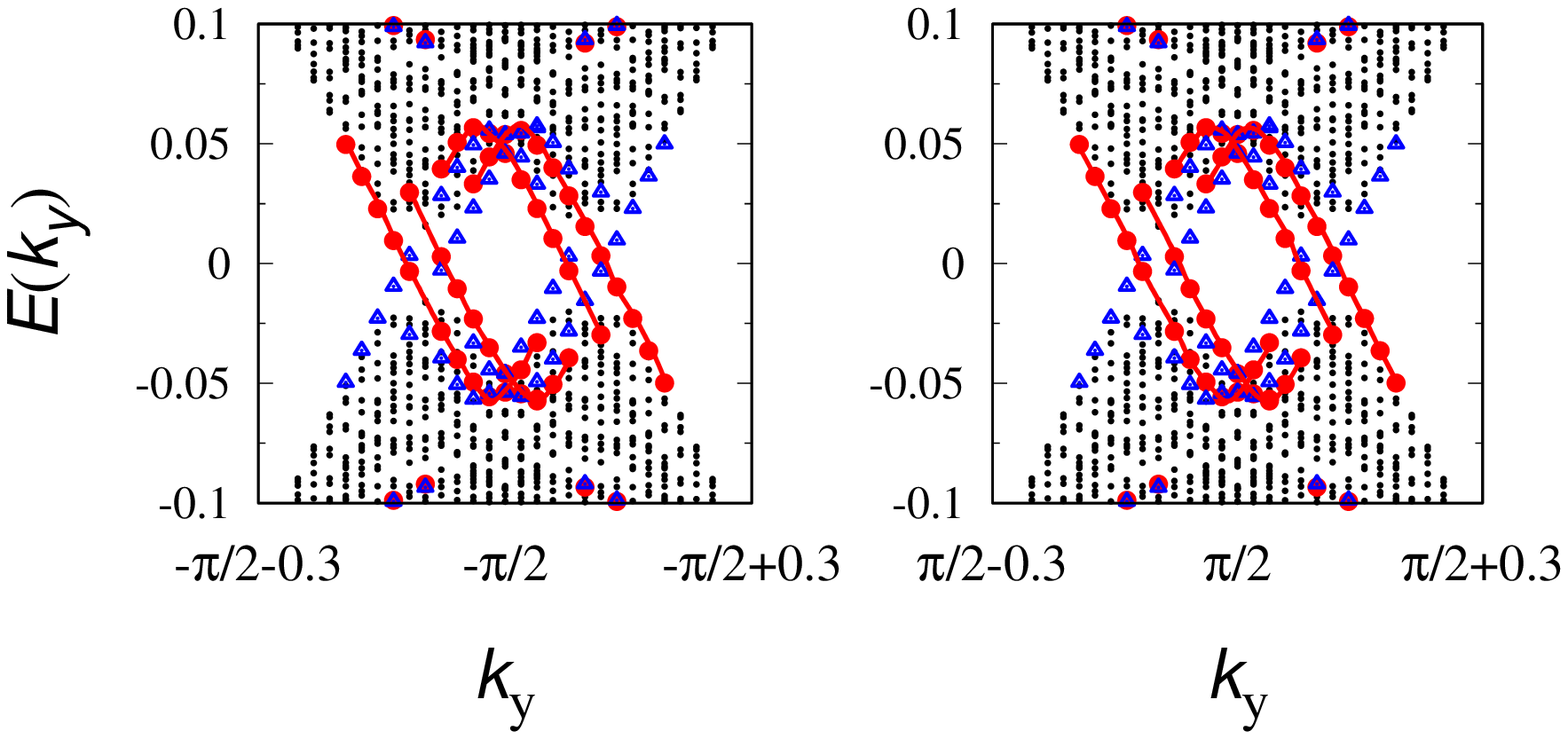}
\end{center}
\caption{(Color Online). 
Energy spectrum of the BdG Hamiltonian, $\mathcal{H}_+(\bm{k})$, around $k_y=-\pi/2$ (left panel) and $k_y=\pi/2$ (right panel) which is obtained under the open (periodic) boundary condition for $x$- ($y$-) direction, respectively.
Here, the spectrum of the eigenstates localized around $x=1$ ($x=L$) is represented by the red (blue) symbols, respectively. The lines are to guide the eye.
The data is obtained with the following parameter set:
$t=1.0$, 
$t_{\bot}=0.1$, 
$\alpha=\alpha'=0.3$,
$\Delta_d=\Delta'_d=0.4$,
$c=c'=-0.08$, and $L=300$, where $L$ denotes the number of sites in the $x$-direction.
At this set of parameters, the system is characterized by $(\nu_\mathrm{M},\nu_\mathrm{tot})=(8,0)$.
Eight Majorana modes are observed around each edge in the system.
}
\label{fig: OBC}
\end{figure}

%%%%%%%%%%%%%%%%%%%%%%%%%
We focus on the region of weak magnetic fields where the topological invariants are $(\nu_+,\nu_-)=(8,-8)$,  predicting eight pairs of helical edge modes. 
To confirm the bulk-edge correspondence, we plot the energy spectrum of $\mathcal{H}_+(\bm{k})$ under the open (periodic) boundary condition in the $x$- ($y$-) direction, respectively (see Fig.~\ref{fig: OBC}). 
%\blue{In this figure, we can find eight Majorana modes localized around each edge, which indicates that the system, \blue{$\mathcal{H}(\bm{k})$}, shows eight-pairs of helical Majorana modes.}
In this figure, we can find that eight Majorana modes localized around $x=1$ ($x=L$) propagate to left (right), respectively. This indicates that the total system, $\mathcal{H}(\bm{k})$, hosts eight pairs of helical Majorana modes.

The above results at the non-interacting level indicate that a topological crystalline superconducting phase with $(\nu_\mathrm{M},\nu_\mathrm{tot})=(8,0)$ emerges in the superlattice with quad-layer of $\mathrm{CeCoIn_5}$. 
%Accordingly, eight pairs of helical Majorana modes appear around the boundary at the non-interacting level.
%%%%%%%%%%%%%%%%%%%%%%%%%
%%%%%%%%%%%%%%%%%%%%%%%%%

%%%%%%%%%%%%%%%%%%%%%%%%%
\textit{- Reduction of topological classification-}
%%%%%%%%%%%%%%%%%%%%%%%%%
So far, we have seen that the system has eight pairs of helical Majorana modes in weak magnetic fields.
Now, by making use of a bosonization approach\cite{Lu_CS_2011,Levin_CS_2012,Hsieh_CS_CPT_2014,Isobe_Fu2015,Yoshida2015}, we analyze the symmetry protection of gapless edge modes in the presence of two-body interactions which are important for the $\mathrm{CeCoIn}_5$/$\mathrm{YbCoIn}_5$. 
%Analysis of symmetry protection elucidates whether the system is topologically trivial or not. 
Our analysis summarized below evidences that these gapless modes are no longer protected by the symmetry. Namely, the topological phase with $(\nu_\mathrm{M},\nu_\mathrm{tot})=(8,0)$ becomes topologically trivial in the presence of two-body interactions. 
In Sec.~III of Ref.~\onlinecite{supp}, we also show that gapless edge modes are protected by the symmetry in the case of two and four pairs of helical Majorana modes. From these results we end up with the classification $\mathbb{Z}\oplus\mathbb{Z}_8$.
We address the above symmetry protection in the following three steps: 
First we show an effective model of gapless edge modes;
We next apply a bosonization scheme to the effective one-dimensional system of edge modes in order to treat one- and two-body interactions on an equal footing;  We finally define the criteria for symmetry protection, and elucidate the fate of Mjorana modes in connection with the symmetry protection in the presence of interactions.

Let us start with an effective model of gapless edge modes, which 
is given for $(\nu_\mathrm{M},\nu_\mathrm{tot})=(8,0)$  as,
%%%%%%%%%%%%%%%%%%
\begin{eqnarray}
H^{edge}&=& \sum_{\alpha,\lambda} \int dx \;\mathrm{sgn}(\lambda)
\eta_{\alpha\lambda}(x) (-iv\partial_x) \eta_{\alpha\lambda}(x),
\label{eq: edge_free}
\end{eqnarray}
%%%%%%%%%%%%%%%%%%
where $\eta_{\alpha\lambda}$ denotes the Majorana operator of a state with $\alpha=1,\cdots,8$ and the eigenvalue of $\mathcal{R}$, $\lambda=+,-$. $v$ denotes the velocity of the Majorana modes.
Under the reflection, these Majorana modes are transformed as 
$\left( \eta_{\alpha+}(x), \eta_{\alpha-}(x)\right) \to \left( -\eta_{\alpha+}(x), \eta_{\alpha-}(x)\right)$. By applying the operator $\hat{P}_f:=(-1)^{N_f}$, the Majorana modes are transformed as $\left( \eta_{\alpha+}(x), \eta_{\alpha-}(x)\right) \to \left( -\eta_{\alpha+}(x),-\eta_{\alpha-}(x)\right)$, where $N_f$ denotes the operator for total number of fermions (see Sec.~IIC and III of Ref.~\onlinecite{supp}).
Here, we note that the symmetry of the fermion number parity and the reflection symmetry are relevant for the symmetry protection of edge modes in the correlated system.
This is because the particle-hole symmetry of the BdG Hamiltonian changes to the symmetry of fermion number parity $\hat{P}_f:=(-1)^{N_f}$ in many-body systems.

We here apply a bosonization scheme to these modes in order to take into account interaction effects. (For details see Sec.~IIIA of Ref.~\onlinecite{supp}.) As a first step, we rewrite two Majorana modes with one complex fermion as follows: 
$\eta_{2\alpha'-1,\lambda}(x)=[f_{\alpha'\lambda}(x)+f^\dagger_{\alpha'\lambda}(x)]/\sqrt{2}$, and $\eta_{2\alpha',\lambda}(x)=[f_{\alpha'\lambda}(x)-f^\dagger_{\alpha'\lambda}(x)]/(\sqrt{2}i)$ with $\alpha'=1,\cdots,4$. 
Transformation law of these operators is the same as that of the Majorana fermions.  Introducing bosonic fields,
%%%%%%%%%%%%%%%%%%
$f_{I}(x)=\frac{1}{\sqrt{2\pi\alpha_0}}\kappa_{I}e^{i\phi_{I}(x)}$,
%%%%%%%%%%%%%%%%%%
the effective action is written as 
%%%%%%%%%%%%%%%%%%
\begin{eqnarray}
\mathcal{L}_{edge}
&=& \int \frac{d\tau dx}{4\pi} \left[ K_{IJ} \partial_\tau \phi_I(x) \partial_x \phi_J(x) \right.  \nonumber \\
&&\quad \quad \quad \quad \quad \left. 
-V_{IJ} \partial_x \phi_I(x) \partial_x \phi_J(x) \right], 
\label{eq: boson_action}
\end{eqnarray}
%%%%%%%%%%%%%%%%%%
with  $K=\sigma^z\otimes \1_{4\times4}$ and $ V=2v\1_{8\times8}$.
Here, $\phi_I(x)$ [$f_{I}(x)$] is a scalar bosonic [fermionic] field and denotes the $I$-th component of the vector fields $\bm{\phi}:=(\phi_{1+},\phi_{1-},\cdots,\phi_{N+},\phi_{N-})^T$ [$\bm{f}:=(f_{1+},f_{1-},\cdots,f_{N+},f_{N-})^T$], respectively.
$\kappa$'s denote the Klein factor, and $\alpha_0$ is a cutoff parameter.

Now, we define the criteria for the symmetry protection of edge modes (see also Sec.~IIIB1 of Ref.~\onlinecite{supp}). If any backscattering term is prohibited from gapping out edge modes by the symmetry, then the edge modes are protected by the symmetry, otherwise the edge modes are not protected. 
Here we formulate (i) backscattering term in terms of bosonic fields and (ii) transformation law of bosonic fields to check whether symmetry is preserved or not. (i) Backscattering terms: Introducing backscattering terms, i.e., mixing right- and left-movers, yields a potential term $\cos(\bm{l}^T\cdot\bm{\phi})$ with $\bm{l}\in \mathbb{Z}^8$, which can pin the field $\bm{l}^T\cdot\bm{\phi}$ to the potential minima and gap out a helical edge mode.
Here, we note that the integer vector $\bm{l}$ satisfies the Haldane's criteria [see Eq.~(29) of Ref.~\onlinecite{supp}]\cite{haldane_nullvector}. 
(ii) Transformation law of bosonic fields: Under the reflection symmetry the bosonic fields are transformed as $\phi_I \to \phi_I +\pi$ for odd $I$, and $\phi_I \to \phi_I$ for even $I$. By applying $\hat{P}_f$, the fields are transformed as $\phi_I \to \phi_I +\pi$ for $I=1,\dots,8$.

Based on the above criteria, we conclude that eight pairs of helical Majorana fermions~(\ref{eq: edge_free}) are no longer protected by the symmetry. 
In other words, the phase labeled by $(\nu_\mathrm{M},\nu_\mathrm{tot})=(8,0)$ is topologically trivial in the presence of two-body interactions. 
%The details are discussed in Sec.~IIIB2 of Ref.~\onlinecite{supp} where we have also found that two and four pairs of helical Majorana fermions cannot be gapped out without breaking the reflection symmetry.
These results indicate that the classification of free fermions, $\mathbb{Z}\oplus\mathbb{Z}$ collapses into $\mathbb{Z}\oplus\mathbb{Z}_8$. The details are discussed in Sec.~IIIB2 of Ref.~\onlinecite{supp}.

Let us now discuss the cases of bi- and tri-layer systems.
In these cases, topological properties of the BdG Hamiltonian is also characterized by $\nu_\mathrm{tot}$ and $\nu_\mathrm{M}$. 
However, the Chern numbers do not predict eight pairs of helical Majorana fermions; $(\nu_\mathrm{M},\nu_\mathrm{tot})$ is given by $(4,0)$ for bi-layer systems, and $(\nu_\mathrm{M},\nu_\mathrm{tot})$  is given by $(1,0)$ for tri-layer systems\cite{superlattice_Yanase16}.  This means that the gapless edge modes localized around the edges of bi- or tri-layer systems cannot be gapped out without symmetry breaking.
Therefore we arrive at the important conclusion: the $\mathrm{CeCoIn}_5/\mathrm{YbCoIn}_5$ superlattice hosts a possible experimental test bed for the reduction of topological classification in free fermion systems, and the minimum number of $\mathrm{CeCoIn}_5$ layers for the reduction is four.

For experimental observation of the gapped edge states, a promising possibility is the scanning tunneling microscopy (STM) measurement. 
The reasons are as follows. (i) This method has been applied to detect the Majorana states emerging at the end of the one-dimensional quantum wires\cite{Mourik_Majorana2012,Nadj-Perge_Majorana2014}. Furthermore, very recently, it becomes possible to carry out the STM measurement for the $\mathrm{CeCoIn_5}/\mathrm{YbCoIn_5}$ superlattice\cite{Matsuda_discussion_2017} where the energy resolution is expected to be high because the STM measurement has already carried out for the bulk $\mathrm{CeCoIn_5}$ with high energy resolution ($\sim75\mu \mathrm{eV}$)\cite{STM_CeCoIn5_2013}. 
(ii) In the superlattice, the interaction arising from antiferromagnetic spin fluctuations is expected to be relevant, so that it contributes to the interactions destroying Majorana edge modes. 
According to our estimation based on the experimental data (STM measurement of pairing potential in the bulk $\mathrm{CeCoIn_5}$\cite{STM_CeCoIn5_2013} and observation of electronic specific heat coefficient\cite{Sidorov_2002}), the interaction arising from antiferromagnetic spin fluctuations is approximately $0.18\mathrm{eV}$. 
Combining this estimation and the numerically obtained wave functions of gapless edge modes under OBC, we conclude that the gap created at the edges is approximately $100\mu \mathrm{eV}$, which is considered to be observable with STM measurement.

Finally we make some comments on the difference between the phase of $(\nu_\mathrm{M},\nu_\mathrm{tot})=(8,0)$ discussed here and an ordinary trivial phase of $(\nu_\mathrm{M},\nu_\mathrm{tot})=(0,0)$.
In the system labeled by $(\nu_\mathrm{M},\nu_\mathrm{tot})=(8,0)$, gapless modes are expected to appear at a dislocation where only the layers $m=2$ and $3$ terminate. This dislocation forms a bi-layer subsystem which is expected to host four pairs of the helical Majorana modes. 

%%%%%%%%%%%%%%%%%%%%%%%%%
\textit{-Summary-}
%%%%%%%%%%%%%%%%%%%%%%%%%
In this paper, we have proposed the $\mathrm{CeCoIn_5}/\mathrm{YbCoIn_5}$ superlattice as a possible experimental test bed for the reduction of the topological classification in free fermion systems. 
Our analysis has elucidated the following results: in the presence of two-body interactions, the classification for the topological crystalline superconductor at the non-interacting level, $\mathbb{Z}\oplus\mathbb{Z}$, collapses into $\mathbb{Z}\oplus\mathbb{Z}_8$, and helical edge modes in the quad-layer system are completely gapped out in the presence of two-body interaction. 
We have demonstrated that the number of $\mathrm{CeCoIn_5}$ layers is essential for detecting the reduction; the minimum number of $\mathrm{CeCoIn_5}$ layers for the reduction is four.
%Besides that, in strong magnetic fields, the system shows sixteen chiral Majorana modes (see Figs.~\ref{fig: ch}(a) and (b)) which are robust even under electron correlation, while there is no gapless mode in weak magnetic fields due to electron correlation. Tuning magnetic fields in this system can help experimental observation of the reduction of topological classification.
Besides that, tuning magnetic fields can help experimental observation of the reduction; In strong magnetic fields, the system shows sixteen chiral Majorana modes (see Figs.~\ref{fig: ch}(a) and (b)) which are robust even under correlations. 

%%%%%%%%%%%%%%%%%%%%%%%%%
\textit{-acknowledgements-}
%%%%%%%%%%%%%%%%%%%%%%%%%
The authors would like to thank Yuji Matsuda for fruitful discussion on the experiments.
This work is partly supported by a Grand-in-Aid for Scientific Research on Innovative Areas (JSPS KAKENHI Grant No.~JP15H05855, No.~JP15H05884, and No.~JP16H00991) and also JSPS KAKENHI (No.~JP25400366 and No.~JP15K05164). The numerical calculations were performed on supercomputer at the ISSP in the University of Tokyo, and the SR16000 at YITP in Kyoto University.

%%%%%%%%%%%%%%%%%%%%%%%%%%%%%%
%\bibliography{superlattice}
%%%%%%%%%%%%%%%%%%%%%%%%%%%%%%


%\appendix
\begin{center}
\textbf{
Supplementary of 
"
Fate of Majorana modes in $\mathrm{CeCoIn}_5$/$\mathrm{YbCoIn}_5$ superlattices
-a test bed for reduction of topological classification-
"
}
\end{center}

%
%
%
%%%%%%%%%%%%%%%%%%%%%%%%%
\section{Crystal symmetry and the pairing potential} \label{ap: GT}
%%%%%%%%%%%%%%%%%%%%%%%%%
Here, by taking into account symmetry of the system, we show that the pairing potential is written as 
%%%%%%%%%%%%%%%%%%
\begin{subequations}
\begin{eqnarray}
\Delta_m     &=& i[\psi(\bm{k})-\bm{d}(\bm{k})\cdot \bm{\sigma}]\sigma^y, \\
\psi(\bm{k}) &=& \Delta_d (\cos k_x- \cos k_y), \\
\bm{d}(\bm{k}) &=& c\left(\sin k_y,\sin k_x,0 \right)^T  \nonumber \\
&&\quad \quad \quad + id \left(\sin k_x,-\sin k_y,0 \right)^T.
\end{eqnarray}
\end{subequations}
%%%%%%%%%%%%%%%%%%

Let us start with the case without magnetic field. In the bulk of $\mathrm{CeCoIn}_5$, the system is invariant under the following symmetry transformations which are generators of $D_{4h}$ group:
$\pi/2$-rotation along $c$-axis ($C_4$); $\pi$-rotation along $a$-axis ($C_2$); $\pi$-rotation along $(1,1,0)$-direction ($C'_2$); reflection whose reflection plane is parallel with $a$- and $b$-axis ($R_z$).
Irreducible representations of pairing potentials for $D_{4h}$ are listed in Table~\ref{table: D4h_symm}.
%In the bulk of $\mathrm{CeCoIn}_5$, the system is invariant under the following symmetry transformations which are generators of $D_{4h}$ group:
%$\pi/2$-rotation along $z$-axis ($C_4$); $\pi$-rotation along $x$-axis $C_2$; $\pi$-rotation along $(1,1,0)$-direction $C'_2$; reflection whose reflection plane is parallel with $x$- and $y$-axis ($R_z$).
%Irreducible representation of $D_{4h}$ is listed in Table~\ref{table: D4h_symm}.
%%%%%%%%%%%
\begin{table}[htb]
\begin{center}
\caption{
Irreducible representation of pairing potential for $D_{4h}$ group\cite{SigristUeda_91}.
}
\begin{center}
\begin{tabular}{c c} \hline \hline
IR        &  $\psi(\bm{k})/\bm{d}(\bm{k})$ \\ \hline       
 $A_{1g}$ & $1$ \\ \hline
 $A_{2g}$ & $k_xk_y(k^2_x-k^2_y)$ \\ \hline
 $B_{1g}$ & $k^2_x-k^2_y$  \\ \hline
 $B_{2g}$ & $k_xk_y$ \\ \hline
 $E_{g}$  & $k_xk_z$, $k_yk_z$ \\ \hline
 $A_{1u}$ & $k_x\bm{x}+k_y\bm{y}$ \\ \hline
 $A_{2u}$ & $k_y\bm{x}-k_x\bm{y}$ \\ \hline
 $B_{1u}$ & $k_x\bm{x}-k_y\bm{y}$ \\ \hline
 $B_{2u}$ & $k_y\bm{x}+k_x\bm{y}$ \\ \hline
 $E_{u}$  & $k_z\bm{x}$,$k_z\bm{y}$; $k_x\bm{z}$,$k_y\bm{z}$ \\ \hline
\end{tabular}
\end{center}
\label{table: D4h_symm}
\end{center}
\end{table}
%%%%%%%%%%%
Here, we have assumed a triplet  $p$-wave paring for simplicity.
For the superlattice systems, local reflection symmetry, i.e., symmetry under applying $R_z$, is broken at each layer of $\mathrm{CeCoIn}_5$\cite{Goh_superlattice12,Shimozawa_superlattice_PRL14,Shimozawa_superlattice_RPP2016}.
In this case the symmetry group changes to $C_{4v}$ of which irreducible representations are listed in Table~\ref{table: C4v_symm}.
%%%%%%%%%%%
\begin{table}[htb]
\begin{center}
\caption{
Irreducible representation of pairing potential for $C_{4v}$ group 
}
\begin{center}
\begin{tabular}{c c} \hline \hline
IR        &  $\psi(\bm{k})/\bm{d}(\bm{k})$ \\ \hline       
 $A_{1}$  & $1$; \\ 
          & $k_y\bm{x}-k_x\bm{y}$ \\ \hline
 $A_{2}$  & $k_xk_y(k^2_x-k^2_y)$; \\ 
          & $k_x\bm{x}+k_y\bm{y}$  \\ \hline
 $B_{1}$  & $k^2_x-k^2_y$;  \\ 
          & $k_y\bm{x}+k_x\bm{y}$ \\ \hline
 $B_{2}$  & $k_xk_y$; \\ 
          & $k_x\bm{x}-k_y\bm{y}$ \\ \hline
 $E$      & $k_xk_z$, $k_yk_z$; \\ 
          & $k_z\bm{x}$,$k_z\bm{y}$; $k_x\bm{z}$,$k_y\bm{z}$ \\ \hline
\end{tabular}
\end{center}
\label{table: C4v_symm}
\end{center}
\end{table}
%%%%%%%%%%%
We note that paring potential is $d_{x^2-y^2}$ in the bulk of $\mathrm{CeCoIn}_5$.
Thus, the superconducting state belongs to the $B_1$ representation, and the $d$-vector of triplet pairing state mixed with $d_{x^2-y^2}$-wave is proportional to $(\sin k_y, \sin k_x,0)^T$. 

Introducing the magnetic field breaks the symmetry under applying $C_2$ and $C'_2$, which reduce the symmetry group from $C_{4v}$ to $C_4$.
The corresponding irreducible representations are listed in Table~\ref{table: C4_symm}.
%%%%%%%%%%%
\begin{table}[htb]
\begin{center}
\caption{
Irreducible representation of pairing potential for $C_{4}$ group 
}
\begin{center}
\begin{tabular}{c c} \hline \hline
IR        &  $\psi(\bm{k})/\bm{d}(\bm{k})$ \\ \hline       
 $A$      & $1$; \\ 
          & $k_xk_y(k^2_x-k^2_y)$; \\ 
          & $k_y\bm{x}-k_x\bm{y}$ \\ 
          & $k_x\bm{x}+k_y\bm{y}$; \\ \hline
 $B$      & $k^2_x-k^2_y$;  \\ 
          & $k_xk_y$; \\ 
          & $k_y\bm{x}+k_x\bm{y}$ \\ 
          & $k_x\bm{x}-k_y\bm{y}$; \\ \hline
 $E$      & $k_xk_z$, $k_yk_z$; \\ 
          & $k_z\bm{x}$,$k_z\bm{y}$; $k_x\bm{z}$,$k_y\bm{z}$ \\ \hline
\end{tabular}
\end{center}
\label{table: C4_symm}
\end{center}
\end{table}
%%%%%%%%%%%
Thus, under the magnetic field, $d$-vector is obtained by superposition of $(\sin k_y, \sin k_x,0)^T$ and $(\sin k_x,-\sin k_y,0)^T$.

Even under the magnetic field, the system is invariant under applying the operator $TR_y$, which fixes the relative phase of pairing potential.
Here, $T$ and $R_y$ denote the time-reversal operator and the reflection operator which maps $(x,y,z)\to(x,-y,z)$, respectively.
%Here, we show that assuming symmetry under applying $TR_y$, the phase factor among $i(\cos k_x -\cos k_y)\sigma^y$, $i(\sin k_y,\sin k_x,0)\cdot \bm{\sigma} \sigma^y$, $i(-\sin k_x,\sin k_y,0)\cdot \bm{\sigma} \sigma^y$ can be fixed.
Under the $TR_y$ operation, a momentum is transformed as $(k_x,k_y)\to (-k_x,k_y)$.  Applying $TR_y$, we have
%%%%%%%%%%%%%%%%%%
\begin{eqnarray}
i(\cos k_x -\cos k_y)\sigma^y &\to& i(\cos k_x -\cos k_y)\sigma^y, \nonumber \\
i(\sin k_y,\sin k_x,0)^T\cdot \bm{\sigma} \sigma^y &\to& i(\sin k_y,\sin k_x,0)^T\cdot \bm{\sigma} \sigma^y, \nonumber \\
i(\sin k_x,-\sin k_y,0)^T\cdot \bm{\sigma} \sigma^y &\to& -i( \sin k_x,-\sin k_y,0)^T\cdot \bm{\sigma} \sigma^y. \nonumber 
\end{eqnarray}
%%%%%%%%%%%%%%%%%%
Thus, the pairing potential is given by 
%%%%%%%%%%%%%%%%%%
\begin{eqnarray}
\Delta_m &=& i[\psi(\bm{k})-\bm{d}(\bm{k})\cdot \bm{\sigma}]\sigma^y,
\end{eqnarray}
%%%%%%%%%%%%%%%%%%
with
%%%%%%%%%%%%%%%%%%
\begin{eqnarray}
\psi_(\bm{k})    &=& \Delta_{d}[\cos(k_x)-\cos(k_y)],\\
\bm{d}(\bm{k})  &=& c(\sin k_y ,\sin k_x ,0)^T +id (\sin k_x, -\sin k_y,0)^T. \nonumber \\ \label{eq: all d} 
\end{eqnarray}
%%%%%%%%%%%%%%%%%%
The second term of Eq.~(\ref{eq: all d}) is considered to just yield quantitative difference for weak magnetic fields and not to alter topological properties. 
Thus, we neglect this term. 
This is because the second term is induced by the magnetic field and much smaller than the first term.

%
%
%
%%%%%%%%%%%%%%%%%%%%%%%%%
\section{Topological properties of the BdG Hamiltonian} \label{ap: BdG}
%%%%%%%%%%%%%%%%%%%%%%%%%

%%%%%%%%%%%%%%%%%%%%%%%%%
\subsection{Relevance of the Hamiltonian to the superlattice}
%%%%%%%%%%%%%%%%%%%%%%%%%
%\blue{
In this section, we explain the relevance of the Hamiltonian of the quad-layer systems to the superlattice systems and also how we have chosen the parameter set. We note that the Hamiltonian for bi- and tri-layer systems can be build up in a similar way.
%}

%\blue{
\textit{-Relevance of the Hamiltonian-}
Concerning the normal part of the Hamiltonian, the interlayer hopping is supposed not to change the number of the Fermi surface. 
Therefore, we employ the dispersion relation of the square lattice with nearest neighbor hopping which is the simplest model describing bulk $\mathrm{CeCoIn_5}$.  
Concerning the pairing potentials, the superconducting phase under the magnetic field is supposed to be a pair-density wave (PDW) phase. 
This is motivated by the following experimental observations. 
The superconducting phase is (i) in the Pauli limit\cite{Onuki_CeCoIn5_Hc2_02} (ii) quasi-two-dimensional in the superlattice 
(iii) affected by strong spin-orbit coupling\cite{Goh_superlattice12,Maruyama_superlattice_12}. 
The Pauli depairing effect affects on the singlet pairing while it does not on the triplet pairing. 
Therefore, the phase of the pairing potentials should be chosen so that the phase of the singlet component changes its sign at each layer, which is characteristic of the PDW phase.
%In the normal part of the Hamiltonian, we employ the simple dispersion relation of the square lattice, which has been used for analysis of bulk $\mathrm{CeCoIn_5}$, because inter-layer hopping supposed not to change the number of Fermi surface. 
%Concerning the pairing potentials, the superconducting phase under the magnetic field is supposed to be a pair-density wave (PDW) phase, which is supported by experimental observation by Goh \textit{et al.}\cite{Goh_superlattice12} that $H_{c2}$ is enhanced in the superlattice. 
%}

%\blue{
\textit{-How to choose the parameter set-} 
To obtain the phase diagram, we chose the parameters as follows. 
Here, we assume that the energy scale of the intra-layer hopping is $3\mathrm{meV}$ by taking into account mass renormalization\cite{Sidorov_2002}.
(i) Pairing potentials: transition temperature of the superconducting phase is approximately $1\mathrm{K}$\cite{Mizukami_CeCoIn5YbCoIn5_11}. With this observation, we set the amplitude of pairing potentials $\Delta_d$ to be $0.05t$. 
(ii) Spin-orbit interactions: first principles calculations predict that the bare anti-symmetric Rashba spin-orbit coupling of heavy fermion is typically 1000K. Taking into account the renormalization factor $\sim1/300$\cite{Sidorov_2002}, we set the parameter $\alpha$ to be $0.3t$.
(iii) Inter-layer hopping: in the $\mathrm{CeCoIn_5/YbCoIn_5}$ superlattice, anomalous angular dependence of the upper critical field $H_{c2}$ is observed experimentally which is attributed to the local inversion symmetry. 
This phenomenon occurs when the inter-layer hopping is smaller than the Rashba spin-orbit coupling. Thus, we have set the parameter $t_\bot=0.1t$.
%}

%%%%%%%%%%%%%%%%%%%%%%%%%
\subsection{Derivation of Eq.~(3)}
%%%%%%%%%%%%%%%%%%%%%%%%%
%In this section, we derive Eq.~(\ref{eq: BdG}).
With the Nambu operator
%$\Psi_{\bm{k}}:=(c_{\bm{k}1\sigma},\cdots, c_{\bm{k}4\sigma},c^\dagger_{-\bm{k}1\sigma},\dots,c^\dagger_{-\bm{k}4\sigma})^T$, 
$\Psi_{\bm{k}}:=\oplus_{\sigma}( c_{\bm{k}1\sigma},\cdots,c_{\bm{k}4\sigma}, c^\dagger_{-\bm{k}1\sigma},\cdots,c^\dagger_{-\bm{k}4\sigma} )^T$
the Hamiltonian is written as $H =\frac{1}{2}\sum_{\bm{k}}\Psi^\dagger_{\bm{k}} \mathcal{H}(\bm{k}) \Psi_{\bm{k}}$. 
We block diagonalize this Hamiltonian with eigenspace of the reflection operator which has the eigenvalue $\lambda i$ ($=i,-i$). 

The reflection operators are written as 
%%%%%%%%%%%%%%%%%%
\begin{eqnarray}
\label{eq: R_mat}
\mathcal{R}&=& i\sigma^z\otimes\tau^0\otimes
\left(
\begin{array}{cccc}
 & & & 1 \\
 & & 1&  \\
 & 1& &  \\
1 & & & 
\end{array}
\right)_L,
\end{eqnarray}
%%%%%%%%%%%%%%%%%%
where $\sigma$ and $\tau$ are the Pauli matrices acting on spin and Nambu space. The $4\times4$-matrix with the subscript $L$ maps the layer $m$ to $5-m$.
%\blue{
We note that the four dimensional matrix with the subscript $L$ is replaced by the two- (three-) dimensional matrix for bi- (tri-) layer systems.
%}

The eigenvectors of the matrix $\mathcal{R}$ are obtained as follows. 
For $\lambda=+$ (the eigenvalues are written as $\lambda i$),
%%%%%%%%%%%%%%%%%%
\begin{subequations}
\begin{eqnarray}
\bm{u}^T_{1+}&=&\frac{1}{\sqrt{2}}(1,0)_\sigma\otimes(1,0)_\tau\otimes(1,0,0,1)_L, \\
\bm{u}^T_{2+}&=&-\frac{1}{\sqrt{2}}(0,1)_\sigma\otimes(1,0)_\tau\otimes(1,0,0,-1)_L, \\
\bm{u}^T_{3+}&=&\frac{1}{\sqrt{2}}(1,0)_\sigma\otimes(1,0)_\tau\otimes(0,1,1,0)_L, \\
\bm{u}^T_{4+}&=&-\frac{1}{\sqrt{2}}(0,1)_\sigma\otimes(1,0)_\tau\otimes(0,1,-1,0)_L, \\
\bm{u}^T_{5+}&=&\frac{1}{\sqrt{2}}(1,0)_\sigma\otimes(0,1)_\tau\otimes(1,0,0,1)_L, \\
\bm{u}^T_{6+}&=&-\frac{1}{\sqrt{2}}(0,1)_\sigma\otimes(0,1)_\tau\otimes(1,0,0,-1)_L, \\
\bm{u}^T_{7+}&=&\frac{1}{\sqrt{2}}(1,0)_\sigma\otimes(0,1)_\tau\otimes(0,1,1,0)_L, \\
\bm{u}^T_{8+}&=&-\frac{1}{\sqrt{2}}(0,1)_\sigma\otimes(0,1)_\tau\otimes(0,1,-1,0)_L,
\end{eqnarray}
\end{subequations}
%%%%%%%%%%%%%%%%%%
where vectors with the subscript $\sigma$ ($\tau$) denote vectors in the spin (Nambu) space, respectively.

For $\lambda=-$ (the eigenvalues are written as $\lambda i$),
%%%%%%%%%%%%%%%%%%
\begin{subequations}
\begin{eqnarray}
\bm{u}^T_{1-}&=&\frac{1}{\sqrt{2}}(0,1)_\sigma\otimes(1,0)_\tau\otimes(1,0,0,1)_L, \\
\bm{u}^T_{2-}&=&\frac{1}{\sqrt{2}}(1,0)_\sigma\otimes(1,0)_\tau\otimes(1,0,0,-1)_L, \\
\bm{u}^T_{3-}&=&\frac{1}{\sqrt{2}}(0,1)_\sigma\otimes(1,0)_\tau\otimes(0,1,1,0)_L, \\
\bm{u}^T_{4-}&=&\frac{1}{\sqrt{2}}(1,0)_\sigma\otimes(1,0)_\tau\otimes(0,1,-1,0)_L, \\
\bm{u}^T_{5-}&=&\frac{1}{\sqrt{2}}(0,1)_\sigma\otimes(0,1)_\tau\otimes(1,0,0,1)_L, \\
\bm{u}^T_{6-}&=&\frac{1}{\sqrt{2}}(1,0)_\sigma\otimes(0,1)_\tau\otimes(1,0,0,-1)_L, \\
\bm{u}^T_{7-}&=&\frac{1}{\sqrt{2}}(0,1)_\sigma\otimes(0,1)_\tau\otimes(0,1,1,0)_L, \\
\bm{u}^T_{8-}&=&\frac{1}{\sqrt{2}}(1,0)_\sigma\otimes(0,1)_\tau\otimes(0,1,-1,0)_L.
\end{eqnarray}
\end{subequations}
%%%%%%%%%%%%%%%%%%

Each element of the block diagonalized BdG Hamiltonian $\mathcal{H}_{\lambda}(\bm{k})$ is written as 
%%%%%%%%%%%%%%%%%%
\begin{eqnarray}
 {} [\mathcal{H}_{\lambda}(\bm{k})]_{ij}&=& \bm{u}^T_{i\lambda} \mathcal{H}(\bm{k}) \bm{u}_{j\lambda}
\end{eqnarray}
%%%%%%%%%%%%%%%%%%
with $i,j=1,\cdots,8$.
%Thus, we endup with Eq.~(\ref{eq: BdG}).
Thus, we end up with the following block diagonalized Hamiltonian,
%%%%%%%%%%%%%%%%%%
\begin{subequations}
\label{eq: h_Delta}
\begin{eqnarray}
\mathcal{H}_{\lambda}(\bm{k})&=& 
\left(
\begin{array}{cc}
\hat{h}_\lambda(\bm{k}) &  \hat{\Delta} (\bm{k}) \\
\hat{\Delta}^\dagger(\bm{k}) & -\hat{h}^T_\lambda(-\bm{k})
\end{array}
\right),
\end{eqnarray}
with
\begin{widetext}
%%%%%%%%%%%%%%%%%%
\begin{eqnarray}
\hat{h}_\lambda(\bm{k}) &=&
\left(
\begin{array}{cccc}
\xi(\bm{k})-\mu_b H & -\alpha k_+ &  t_\bot & 0 \\
-\alpha k_- & \xi(\bm{k})+\mu_b H &  0 & t_\bot \\
 t_\bot & 0 & \xi(\bm{k}) -\mu_b H + \mathrm{sgn}(\lambda)t_\bot  & -\alpha' k_+  \\
 0 & t_\bot & -\alpha' k_- & \xi(\bm{k}) + \mu_b H - \mathrm{sgn}(\lambda)t_\bot
\end{array}
\right),
\\
\hat{\Delta} (\bm{k}) &=&
\left(
\begin{array}{cccc}
d_x(\bm{k})-id_y(\bm{k}) & \psi(\bm{k})  & 0 & 0  \\
-\psi(\bm{k}) & -d_x(\bm{k})-id_y(\bm{k})& 0 & 0  \\
0 & 0 & d'_x(\bm{k})-id'_y(\bm{k}) & \psi'(\bm{k})    \\
0 & 0 &-\psi'(\bm{k}) & -d'_x(\bm{k})-id'_y(\bm{k})   
\end{array}
\right),
\end{eqnarray}
%%%%%%%%%%%%%%%%%%
\end{widetext}
\end{subequations}
%%%%%%%%%%%%%%%%%%
where $k_{\pm}=\sin(k_y)\pm i\sin(k_x)$, and $\mathrm{sgn}(\lambda=\pm)=\pm1$.

%%%%%%%%%%%%%%%%%%
\subsection{topological class}\label{ap: BdG_topo_class}
%%%%%%%%%%%%%%%%%%
In this section, we show that the topological class of the BdG Hamiltonian is class D.
The Hamiltonian $\mathcal{H}(\bm{k})$ respects the reflection and the particle-hole symmetry. Namely, it satisfies
%%%%%%%%%%%%%%%%%%
\begin{eqnarray}
\mathcal{R} \mathcal{H}(\bm{k}) \mathcal{R}^{-1} &=& \mathcal{H}(\bm{k}), \\
\mathcal{C} \mathcal{H}(-\bm{k}) \mathcal{C}^{-1} &=& -\mathcal{H}(\bm{k}).
\end{eqnarray}
%%%%%%%%%%%%%%%%%%
where $\mathcal{C}:=\tau^x\mathcal{K}$ ($\mathcal{C}^2=\1$) denotes the operator of the particle-hole transformation.
%\tau^x is the Pauli matrix acting on the Nambu space and the operator taking complex conjugation. 
%$\mathcal{R}$ ($\mathcal{R}^2=-\1$) denotes the reflection.
$\mathcal{R}$ and $\mathcal{C}$ anti-commute, $\{\mathcal{R},\mathcal{C}\}=0$.
Thus, the block diagonalized Hamiltonian $\mathcal{H}_{\lambda}(\bm{k})$ also respects the particle-hole symmetry, and the topological class is D.
The above results mean that the block diagonalized Hamiltonian $\mathcal{H}_{\lambda}(\bm{k})$ is characterized by Chern number $\nu_\pm$ which is defined in Eq.~(\ref{eq ch_def}). 
In other words, the BdG Hamiltonian $\mathcal{H}(\bm{k})$ is characterized by the total Chern number and the mirror Chern number defined in Eq.~(4). 
We note that the mirror Chern number takes multiple of $1/2$, which is just due to our convention.

We note that for $\mu_BH=0$, the system is invariant under the time-reversal symmetry. Namely,
%%%%%%%%%%%%%%%%%%
\begin{eqnarray}
\mathcal{T}
\mathcal{H}(-\bm{k})
\mathcal{T}^{-1}
&=&
\mathcal{H}(\bm{k}),
\end{eqnarray}
%%%%%%%%%%%%%%%%%%
with $\mathcal{T}:=i\sigma^y \mathcal{K}$. However, the block diagonalized Hamiltonian $\mathcal{H}_{\lambda}(\bm{k})$ does not respect the time-reversal symmetry because $\mathcal{T}\bm{u}_{i\pm}=\mp\bm{u}_{i\mp}$ holds. 
Therefore, the mirror Chern number may take a finite value.

We finish this part by noting that the overall phase in Eq.~(\ref{eq: R_mat}) is just for convention. 
We can also choose the following reflection operator instead of the operator in Eq.~(\ref{eq: R_mat}):
%%%%%%%%%%%%%%%%%%
\begin{eqnarray}
\label{eq: R_mat'}
\mathcal{R}'&=&
-\sigma^z \otimes \tau^0 \otimes 
\left(
\begin{array}{cccc}
 & & & 1 \\
 & &1 &  \\
 &1 & &  \\
1 & & & 
\end{array}
\right),
\end{eqnarray}
%%%%%%%%%%%%%%%%%%
which satisfies $\mathcal{R}'^2=\1$.
This does not change the results in the non-interacting system. 
Indeed, we can check the following facts: 
The BdG Hamiltonian commutes with $\mathcal{R}'$; 
The block diagonalized Hamiltonian in the eigenspace of $\mathcal{R}'$ is given by Eqs.~(\ref{eq: h_Delta});
The topological class of the block diagonalized Hamiltonians is class D because eigenvalues of $\mathcal{R}'$ is $\pm1$, and the commutation relation $[\mathcal{R}',\mathcal{C}]=0$ holds.
Therefore, the topological properties of the BdG Hamiltonian are characterized by Chern numbers for each sector $\nu_\pm$.

%%%%%%%%%%%%%%%%%%
\subsection{topological invariants}
%%%%%%%%%%%%%%%%%%
In the previous section, we have seen that the topological class of the two-dimensional Hamiltonian $\mathcal{H}_{\lambda}(\bm{k})$ is class D.
Therefore, the Chern number of each subspace characterizes the topological structure of the BdG Hamiltonian of the two-dimensional system, or equivalently, the total Chern number and the mirror Chern number [Eq.~(4) in the main text] characterize the topology,
%%%%%%%%%%%%%%%%%%
\begin{subequations}
\label{eq ch_def}
\begin{eqnarray}
\nu_{\pm}&=& \frac{1}{2\pi i} \int_{BZ} \!\!\!\!  d^2\bm{k} \;F^\lambda_{12}(\bm{k}),
\end{eqnarray}
%%%%%%%%%%%%%%%%%%
where $F^\lambda_{12}(\bm{k})$ denotes Berry curvature,
%%%%%%%%%%%%%%%%%%
\begin{eqnarray}
F^\lambda_{12}(\bm{k})&=& \partial_1 A^\lambda_2(\bm{k}) - \partial_2 A^\lambda_1(\bm{k}), \\
A^\lambda_\mu(\bm{k})&=& i\langle n_\lambda (\bm{k})| \partial_\mu |n_\lambda (\bm{k}) \rangle.
\end{eqnarray}
\end{subequations}
%%%%%%%%%%%%%%%%%%
Here, $|n_\lambda (\bm{k}) \rangle$ denotes a normalized wave function of the $n$-th Bloch state of $\mathcal{H}_\lambda(\bm{k})$; 
$\mathcal{H}_\lambda(\bm{k}) |n_\lambda (\bm{k}) \rangle =E_{n\lambda} |n_\lambda (\bm{k}) \rangle $.

Here, a technical comment is in order on the computation of the Chern number $\nu_{\pm}$.
The computation of the Chern numbers is numerically difficult because of the following reasons.
Firstly, dimension of the Hamiltonian matrix $\mathcal{H}_{\lambda}(\bm{k})$ is eight, $\mathrm{dim}\mathcal{H}_{\lambda}(\bm{k})=8$, which makes the calculation of the Chern number heavy. 
Secondly, the bulk gap for the above parameter set is so small that the Chern number cannot be easily computed with sufficient accuracy by using Eq.~(\ref{eq: nu_gf}). 
Therefore, instead of direct computation of the Chern number with Eq.~(\ref{eq ch_def}), we use the method proposed in Ref.~\onlinecite{Fukui_Hatsugai_05}. 
Following this method, we discretize the momentum space, and computes the Berry curvature from a U(1) link variable for each discretized patches. 
%\blue{
This formula is applicable not only to the quad-layer system but also to other multiple layer systems (e.g., bi- and tri-layer systems).
%}

We note that the Chern number can be rewritten with the single-particle Green's function;
%%%%%%%%%%%%%%%%%%
\begin{eqnarray}
\label{eq: nu_gf}
 \nu_{\lambda}&=& \epsilon^{\mu\nu\rho} \int \frac{d\omega d^2\bm{k}}{24\pi^2} \mathrm{tr}[G^{-1}_{\lambda}(k)\partial_\mu G_{\lambda}(k) \nonumber \\
 && \quad \quad \times G^{-1}_{\lambda}(k)\partial_\nu G_{\lambda}(k)   G^{-1}_{\lambda}(k)\partial_\rho G_{\lambda}(k)],
\end{eqnarray}
%%%%%%%%%%%%%%%%%%
where $\epsilon^{\mu\nu\rho}$ is the anti-symmetric tensor ($\epsilon^{012}=1$, $\mu,\nu,\rho=0,1,2$). $\bm{\partial}:=(\partial_\omega,\partial_{k_x},\partial_{k_y})$ and $\bm{k}:=(k_x,k_y)$. Here, $G_{\lambda}(k)$ with $k:=(i\omega,\bm{k})$ is a matrix whose elements are the single-particle Green's function for the eigenspace labeled by $\lambda=\pm$.
The Chern number in Eq.~(\ref{eq: nu_gf}) is well-defined as long as the Green's function is non-singular; $\mathrm{det}G_{\lambda}(k)\neq 0$ and $\mathrm{det} G^{-1}_{\lambda}(k)\neq 0$ for ${}^\forall k \in (i\omega,\bm{k})$.
Thus, the Chern number $\nu_{\pm}$ can take a quantized value even in the correlated systems.

%Thus, the topology of the original Hamiltonian $\mathcal{H}(\bm{k})$ is characterized by the total Chern number $\mathcal{\nu}_\mathrm{tot}$ and the mirror Chern number $\nu_M$ in Eq.~(\ref{eq: nu_tot}). \green{(***Which is better labeling $n_M$ or $n_\pm$***?)}
%
%
%
%%%%%%%%%%%%%%%%%%%%%%%%%
\section{Analysis of gapless edge modes in the presence of correlations} \label{ap: bosonization}
%%%%%%%%%%%%%%%%%%%%%%%%%
In this section, we show the following results by taking into account one- and two-body interaction.
Two and four pairs of helical Majorana modes are protected by the reflection symmetry and the symmetry of fermion number parity, while eight pairs of helical Majorana modes are not.
From this result, we end up with the classification results $\mathbb{Z}\oplus\mathbb{Z}_8$ because (i) chiral modes cannot be gapped out, and (ii) odd number of helical Majorana modes are considered to be stable against interactions\cite{YaoRyu_Z_to_Z8_2013,Ryu_Z_to_Z8_2013,Hsieh_CS_CPT_2014,You_Cenke2014,Morimoto_2015}. 
We note that the particle-hole symmetry of the BdG Hamiltonian Eq.~(3) is replaced by the symmetry of fermion number parity in the presence of electron correlations. 
We denote the operator for the fermion number parity as $\hat{P}_f:=(-1)^{N_f}$ with the number operator of total fermions $N_f$.
In the following, we discuss how we end up with the classification results.

Consider a one-dimensional subsystem of the topological crystalline superconductors having the  $2N$ $(=2,4,8)$  helical edge modes.
The Hamiltonian is written as
%%%%%%%%%%%%%%%%%
\begin{eqnarray}
H^{edge}&=& \sum^{2N}_{\alpha=1}\int dx \;\mathrm{sgn}(\lambda) \eta_{\alpha\lambda}(x)(-i v \partial_x)\eta_{\alpha\lambda}(x), 
\label{eq: free_Majorana_ap}
\end{eqnarray}
%%%%%%%%%%%%%%%%%%
where $\eta_{\alpha\lambda}$ denotes the Majorana operator for the state with $\alpha=1,2,\cdots,2N$ and $\lambda=+,-$. $\mathrm{sgn}(\lambda)$ takes $1$ ($-1$) for $\lambda=+$ ($-$), respectively. 
$v$ denotes the velocity of the edge modes. 
Under each symmetry transformation, these Majorana modes are transformed as follows:
%%%%%%%%%%%%%%%%%%
\begin{subequations}
\begin{eqnarray}
\hat{R} \eta_{\alpha\lambda}(x) \hat{R}^{-1} &=& -\mathrm{sgn}(\lambda) \eta_{\alpha\lambda}, \\
\hat{P}_f \eta_{\alpha\lambda}(x) \hat{P}_f^{-1} &=& -\eta_{\alpha\lambda}.
\end{eqnarray}
\end{subequations}
%%%%%%%%%%%%%%%%%%

As we note in Sec.~\ref{ap: BdG_topo_class}, the overall phase in Eq.~(\ref{eq: R_mat}) is just for convention. 
Here, we choose the matrix in Eq.~(\ref{eq: R_mat'}) as the reflection operator. This does not change the topological classification for free fermions, $\mathbb{Z}\oplus\mathbb{Z}$.
In this case, the reflection operator $\hat{R}$ is given by
%%%%%%%%%%%%%%%%%
\begin{eqnarray}
\label{eq R_operator}
\hat{R}&=& e^{-i\frac{\pi}{2}N} e^{-i\pi S^z}P,
\end{eqnarray}
%%%%%%%%%%%%%%%%%%
where $P$ denotes the operator satisfying $Pc_{im\sigma}P^{-1}=c_{i5-m\sigma}$. 
Multiplying the overall phase $-i$ yelds Eq.~(\ref{eq: R_mat}). 
This additional overall phase changes the commutation relation. 

%Defining $\eta_{2\alpha'-1,\lambda}(x)=[f_{\alpha'\lambda}(x)+f^\dagger_{\alpha'\lambda}]/\sqrt{2}$ and $\eta_{2\alpha,\lambda}(x)=(f_{\alpha'\lambda}(x)-f^\dagger_{\alpha'\lambda})/(\sqrt{2}i)$ with $\alpha=1,\cdots,N$,

%%%%%%%%%%%%%%%%%%%%%%%%%
\subsection{bosonization of edge modes} \label{ap: bosonization_of_Majorana}
%%%%%%%%%%%%%%%%%%%%%%%%%
In order to discuss the interaction effects, we bosonize the one-dimensional model. As a first step, we rewrite the Hamiltonian~(\ref{eq: free_Majorana_ap}) with complex fermions to bosonize the model.
By introducing the complex fermions,
%%%%%%%%%%%%%%%%%%
\begin{subequations}
\begin{eqnarray}
\eta_{2\alpha'-1,\lambda}(x)
&=&
\frac{1}{\sqrt{2}} [f_{\alpha'\lambda}(x)+f^\dagger_{\alpha'\lambda}],\\
\eta_{2\alpha',\lambda}(x)
&=&
\frac{1}{\sqrt{2}i} [f_{\alpha'\lambda}(x)-f^\dagger_{\alpha'\lambda}],
\end{eqnarray}
\end{subequations}
%%%%%%%%%%%%%%%%%%
with $\alpha'=1,\cdots,N$, Eq~(\ref{eq: free_Majorana_ap}) is written as 
%%%%%%%%%%%%%%%%%%
\begin{eqnarray}
H^{edge}&=& \sum_{\alpha',\lambda} v \int dx \; \mathrm{sgn}(\lambda)
[\{i\partial_xf_{\alpha'\lambda}(x)\}f^\dagger_{\alpha'\lambda}(x)\nonumber \\
&&\quad \quad \quad \quad \quad+f^\dagger_{\alpha'\lambda}(x)(-i\partial_x)f_{\alpha'\lambda}(x)],\nonumber \\
&=&
\sum_{\alpha',\lambda}
\int dx 
[2 v \, \mathrm{sgn}(\lambda)
f^\dagger_{\alpha'\lambda}(x)(-i\partial_x)f_{\alpha'\lambda}(x)], \nonumber \\
\end{eqnarray}
%%%%%%%%%%%%%%%%%%
where $\alpha'$ runs from $1$ to $N$ in the summation.
%Applying the bosonization scheme, we can examine the instability/stability of edge modes against the two-body interactions because the edge modes are described by the one-dimensional effective model.
%Let us discuss the instability/stability of edge modes against the interaction. Here we restrict ourselves to the case of one- and two-body interactions. As the edge modes are described by the one-dimensional effective model, we can treat the interaction effects with the bosonization approach.

Now, we bosonize the model written with the complex fermions.
Introducing bosonic fields,
%%%%%%%%%%%%%%%%%%
\begin{eqnarray}
f_{I}(x)&=&\frac{1}{\sqrt{2\pi\alpha}}\kappa_{I}e^{i\phi_I(x)},
\end{eqnarray}
%%%%%%%%%%%%%%%%%%
the effective action is written as 
%%%%%%%%%%%%%%%%%%
\begin{subequations}
\label{eq: boson_action}
\begin{eqnarray}
\label{eq: boson_action_a}
\mathcal{L}_{edge}
&=& \int \frac{d\tau dx}{4\pi} \left[ K_{IJ} \partial_\tau \phi_I(x) \partial_x \phi_J(x) \right.  \nonumber \\
&&\quad \quad \quad \quad \quad \left. -V_{IJ} \partial_x \phi_I(x) \partial_x \phi_J(x) \right], 
\end{eqnarray}
with
\begin{eqnarray}
 K=\sigma^z\otimes \1_{N\times N},     &\quad& V =2v\1_{2N\times 2N}.
\end{eqnarray}
\end{subequations}
%%%%%%%%%%%%%%%%%%
%Here $\phi_a$ is $a$-th component of the bosonic fields $\bm{\phi}:=(\phi_{1+},\phi_{1-},\cdots,\phi_{N+},\phi_{N-})$, $f_{a}$ denotes $a$-th component of the vector $(f_{1+},f_{1-},\cdots,f_{N+},f_{N-})$, and $\kappa$'s denote the Klein factor. $\alpha_0$ denotes cutoff parameter.
%Here, $\phi_I(x)$ [$f_{I}(x)$] denotes $I$-th component of the bosonic fields $\bm{\phi}:=(\phi_{1+},\phi_{1-},\cdots,\phi_{N+},\phi_{N-})$ [the fermionc \green{fields} $\bm{f}:=(f_{1+},f_{1-},\cdots,f_{N+},f_{N-})$]. $\kappa$'s denote the Klein factor. $\alpha_0$ denotes cutoff parameter.
Here, $\phi_I(x)$ [$f_{I}(x)$] is a scalar bosonic (fermionic) field and denotes $I$-th component of the vector fields $\bm{\phi}:=(\phi_{1+},\phi_{1-},\cdots,\phi_{N+},\phi_{N-})^T$ [$\bm{f}:=(f_{1+},f_{1-},\cdots,f_{N+},f_{N-})^T$], respectively. 
$\kappa$'s denote the Klein factor and $\alpha_0$ is a cutoff parameter.

We note that the commutation relations of $\phi$'s are encapsulated in the first term of Eq.~(\ref{eq: boson_action_a}), which are written as 
%%%%%%%%%%%%%%%%%%
\begin{eqnarray}
{}[\phi_I(x),\phi_J(x')]&=& \pi i [K^{-1}]_{IJ}\mathrm{sgn}(x-x'),
\end{eqnarray}
%%%%%%%%%%%%%%%%%%
where $\mathrm{sgn}(x)$ equals $1$, $0$, and $-1$ for $x>0$, $x=0$, and $x<0$, respectively. 
Introducing backscattering terms yields a $\cos$-term. If a bosonic field is pinned at the potential minimum, one pair of the helical edge modes described by complex fermions (i.e., two pairs of Majorana helical edge modes) are gapped out.

Under the reflection, the bosonic fields are transformed as 
%%%%%%%%%%%%%%%%%%
\begin{subequations}
\label{eq: M_phi_M}
\begin{eqnarray}
\hat{R}
\bm{\phi}
\hat{R}^{-1}
&=&
\bm{\phi}
+\delta \bm{\phi_R},
\end{eqnarray}
with
\begin{eqnarray}
\delta \bm{\phi_R}
&=&
\pi(1,0,1,0,\cdots,1,0)^T.
\end{eqnarray}
\end{subequations}
%%%%%%%%%%%%%%%%%%
Namely,  applying the operator $\hat{R}$ yields the phase shift $\pi$ only for $\phi_I$ with odd $I\in(1,\dots,N)$.
Under $\hat{P}_f$, the bosonic fields are transformed as 
%%%%%%%%%%%%%%%%%%
\label{eq: P_phi_P}
\begin{eqnarray}
\hat{P}_f
\bm{\phi}
\hat{P}^{-1}_f
&=&
\bm{\phi}
+\pi (1,1,\cdots,1,1)^T.
\end{eqnarray}
%%%%%%%%%%%%%%%%%%

%%%%%%%%%%%%%%%%%%%%%%%%%
\subsection{symmetry protection of edge modes} \label{ap: symmetry protection}
%%%%%%%%%%%%%%%%%%%%%%%%%
%%%%%%%%%%%%%%%%%%%%%%%%%
\subsubsection{criteria for symmetry protection} \label{ap: criteria for SP}
%%%%%%%%%%%%%%%%%%%%%%%%%
Symmetry protection of edge modes against one- and two-body interactions can be discussed as follows.
One pair of the helical edge modes of complex fermion are gapped out when a bosonic field is pinned at the potential minimum of backscattering terms. 
If  all of the gapless modes are gapped out without symmetry breaking by interactions, the gapless modes are not protected by the symmetry. Otherwise the gapless modes are protected by symmetry. 
In the following, we discuss how to check the symmetry breaking and thus elucidate the condition for the backscattering term to gap out all of the edge modes.

Firstly, we discuss the symmetry breaking. 
The reflection symmetry breaking of the Hamiltonian can be checked by applying the operator $\hat{R}$ to the backscattering term. (We can discuss the fermion number parity breaking in a similar way.)
In addition, one should check the absence of spontaneous symmetry breaking.
Let us start with the symmetry of the Hamiltonian. A backscattering term can be written as
%%%%%%%%%%%%%%%%%%
\begin{eqnarray}
\mathcal{L}_\mathrm{int}&=& U_1 \int dx \cos(\phi_1-\phi_2),
\end{eqnarray}
%%%%%%%%%%%%%%%%%%
where $U_1$ is a real number.  
Applying $\hat{R}$ transforms the interaction term as follows:
%%%%%%%%%%%%%%%%%%
\begin{eqnarray}
\hat{R} \mathcal{L}_\mathrm{int} \hat{R}^{-1}&=&-U_1\int dx \cos(\phi_1-\phi_2).
\end{eqnarray}
%%%%%%%%%%%%%%%%%%
Thus, this interaction term breaks the symmetry. 
To check spontaneous symmetry breaking in the bosonization approach, consider the interaction term
%%%%%%%%%%%%%%%%%%
\begin{eqnarray}
\mathcal{L}_\mathrm{int}&=& U_1 \int dx \cos(2\phi_1-2\phi_2),
\end{eqnarray}
%%%%%%%%%%%%%%%%%%
which is invariant under applying $\hat{R}$. If the bosonic field is pinned at the potential minimum (i.e., $\langle 2\phi_1-2\phi_2\rangle=const.$) gapless edge modes are gapped out. 
In this case, however, a more primitive field $\langle \phi_1-\phi_2\rangle=const.$ (we call this elementary bosonic field) is not invariant, whereby the symmetry is spontaneously broken.

Secondly, we note that an arbitrary set of back scattering terms  $\{\cos(\bm{l}^T_1\cdot \bm{\phi}),\cdots,\cos(\bm{l}^T_i\cdot \bm{\phi}),\cdots\}$ pinning fields $\bm{l}^T_i\cdot \bm{\phi}$ ($i=1,2\cdots$) satisfies the following Haldane criteria:
%%%%%%%%%%%%%%%%%%
\begin{eqnarray}
\bm{l}^T_iK^{-1}\bm{l}_j&=&0,
\label{eq: Haldane_criteria}
\end{eqnarray}
%%%%%%%%%%%%%%%%%%
where $\bm{l}$'s are integer vectors since the back scattering terms are generated by creation or annihilation operators of the complex fermions. This is because if the fields $\bm{l}^T_i\cdot \bm{\phi}$ are pinned simultaneously the fields must be commutative $[\bm{l}^T_i\cdot \bm{\phi}(x),\bm{l}^T_j\cdot \bm{\phi}(x')]=2i\pi \bm{l}^T_i K^{-1}\bm{l}_j \mathrm{sgn}(x-x')$.

In the following, we show that two and four pairs of helical Majorana modes are protected by the symmetry even in the presence of two-body interactions, and that eight pairs of helical Majorana modes are gapped out without symmetry breaking (i.e., the system characterized by $(\nu_\mathrm{M},\nu_\mathrm{tot})=(8,0)$ is topologically trivial in the presence of the interactions).
We address the stability/instability of gapless modes step by step below.

%%%%%%%%%%%%%%%%%%%%%%%%%
\subsubsection{symmetry protection of two-, four- eight- Majorana modes} \label{ap: calc. for SP}
%%%%%%%%%%%%%%%%%%%%%%%%%
-\textit{two pairs of helical Majorana modes}-
In this case, we can gap out helical edge modes by introducing one of the following two-body interaction terms to the model (\ref{eq: boson_action}):
%%%%%%%%%%%%%%%%%%
\begin{subequations}
\begin{eqnarray}
\mathcal{L}_\mathrm{int}&=&U_1 \int \! dx \cos(2\phi_1+2\phi_2),
\end{eqnarray}
or
\begin{eqnarray}
\mathcal{L}_\mathrm{int}&=&U'_1 \int \! dx   \cos(2\phi_1-2\phi_2),
\end{eqnarray}
\end{subequations}
%%%%%%%%%%%%%%%%%%
where $U_1$ and $U'_1$ are real numbers.
In both cases, the reflection symmetry is spontaneously broken because $\hat{R}(\phi_1\pm\phi_2)\hat{R}^{-1}=(\phi_1\pm\phi_2)+\pi$ holds.
In other words, the gapless edge modes cannot be gapped out without symmetry breaking.
Thus, the gapless edge modes in this case are protected by the symmetry.
%We note that the symmetry under appling $\mathrm{P}_f$ is not broken for $K=\otimes \sigma^z$ because p%inned bosonic fields by the backscattering term always composed of even numbers of the bosonic fields.
%Hence, it is sufficient to discuss the symmetry protection by the reflection symmetry.\\

-\textit{four pairs of helical Majorana modes}-
In this case, we need two $\cos$-terms in order to gap out all of the gapless edge modes.
These interaction terms are given by
%%%%%%%%%%%%%%%%%%
\begin{subequations}
\label{eq: potential_fourpair}
\begin{eqnarray}
\mathcal{L}_\mathrm{int}&=&\int \! dx \left[ U_1 \cos(\bm{l}^T_1\cdot\bm{\phi}) + U_2 \cos(\bm{l}^T_2\cdot\bm{\phi})\right], \nonumber \\
\end{eqnarray}
%%%%%%%%%%%%%%%%%%
where $U$'s are real numbers. $\bm{l}$'s are linear independent integer vectors given by the following set
%%%%%%%%%%%%%%%%%%
\begin{eqnarray}
\{\bm{l}^T_1,\bm{l}^T_2\}&=&\{(1,1|1,1),(1,-1|-1,1)\}, \label{eq: potential_fourpair_1} \\
&& \{(1,-1|1,-1),(1,1|-1,-1)\} \label{eq: potential_fourpair_2}.
\end{eqnarray}
%%%%%%%%%%%%%%%%%%
\end{subequations}
%%%%%%%%%%%%%%%%%%
In both cases, the symmetry is spontaneously broken because the elementary bosonic fields are not invariant under the transformation. 
If the potential terms are given by Eq.~(\ref{eq: potential_fourpair_1}), the elementary bosonic fields $\bm{v}^T_1\cdot\bm{\phi}$ and $\bm{v}^T_2\cdot\bm{\phi}$ are given by
%%%%%%%%%%%%%%%%%%
\begin{eqnarray}
\bm{v}^T_1=(1,0|0,1), &\quad& \bm{v}^T_2=(0,1|1,0).
\end{eqnarray}
%%%%%%%%%%%%%%%%%%
On the other hand, if the potential terms are given by Eq.~(\ref{eq: potential_fourpair_2}), the elementary bosonic fields are obtained by substituting
%%%%%%%%%%%%%%%%%%
\begin{eqnarray}
\bm{v}^T_1=(1,0|0,-1), &\quad& \bm{v}^T_2=(0,1|-1,0).
\end{eqnarray}
%%%%%%%%%%%%%%%%%%
In both cases, the relation $\hat{R}\bm{v}^T_i\cdot\bm{\phi}\hat{R}^{-1}=\bm{v}^T_i\cdot\bm{\phi}+\pi$ ($i=1,2$) holds, and the symmetry is spontaneously broken.

-\textit{eight pairs of helical Majorana modes}-
In this case, we need four $\cos$-terms in order to gap out all of the gapless edge modes.
These interaction terms are given by
%%%%%%%%%%%%%%%%%%
\begin{subequations}
\label{eq: potential_fourpair}
\begin{eqnarray}
\mathcal{L}_\mathrm{int}&=&\sum^4_{i=1} U_i\int dx  \cos(\bm{l}^T_i\cdot\bm{\phi}),
\end{eqnarray}
where $U$'s are real numbers. $\bm{l}$'s are linear independent integer vectors given by the following set
%%%%%%%%%%%%%%%%%%
\begin{eqnarray}
\bm{l}^T_1 &=& (1,0|1,0|0,-1|0,-1), \\
\bm{l}^T_2 &=& (0,1|0,1|-1,0|-1,0), \\
\bm{l}^T_3 &=& (1,1|-1,-1|0,0|0,0), \\
\bm{l}^T_4 &=& (0,0|0,0|1,1|-1,-1). 
\end{eqnarray}
%%%%%%%%%%%%%%%%%%
\end{subequations}
%%%%%%%%%%%%%%%%%%
These interaction terms can gap out all of helical edge modes without symmetry breaking. Thus, the eight pairs of helical Majorana modes are not protected by the symmetry.

We note that the following set of vectors $\bm{l}$ can also gap out all of the helical edge modes.
%%%%%%%%%%%%%%%%%%
\begin{subequations}
\begin{eqnarray}
\bm{l}^T_1 &=& (1,0|1,0|0,-1|0,-1), \\
\bm{l}^T_2 &=& (0,1|0,1|-1,0|-1,0), \\
\bm{l}^T_3 &=& (1,-1|1,-1|0,0|0,0), \\
\bm{l}^T_4 &=& (0,0|0,0|1,-1|1,-1).
\end{eqnarray}
\end{subequations}
%%%%%%%%%%%%%%%%%%

In the above, we have seen that eight pairs of helical Majorana modes are no longer symmetry protected in the presence of two-body interactions, while two and four pairs of helical Majorana modes are protected.
We note that these edge modes are predicted by the mirror Chern number in the bulk. These results demonstrate that six pairs of Majorana modes are protected by the symmetry because a phase with the mirror Chern number $\nu_\mathrm{M}=6(=8-2)$ is topologically equivalent to a phase having two pairs of helical Majorana modes predicted by the mirror Chern number $\nu_\mathrm{M}=-2$ in the bulk.
Besides that, the chiral Majorana modes and odd number of helical Majorana modes are robust against interactions\cite{YaoRyu_Z_to_Z8_2013,Ryu_Z_to_Z8_2013,Hsieh_CS_CPT_2014,You_Cenke2014,Morimoto_2015}.
Therefore, we end up with the classification results, $\mathbb{Z}\oplus\mathbb{Z}_8$ in the presence of two-body interactions.

We finish this part by making a comment on interactions which gap out edge modes.
In real materials, any symmetry-allowed interactions are supposed to exist. Thus, the layers of $\mathrm{CeCoIn}_5$ are expected to host interactions gapping out edge modes. 
For qualitative estimation of interactions, however, one need microscopic analysis based on Hubbard or periodic Anderson model, which is left for the future work.


\begin{thebibliography}{57}%
\makeatletter
\providecommand \@ifxundefined [1]{%
 \@ifx{#1\undefined}
}%
\providecommand \@ifnum [1]{%
 \ifnum #1\expandafter \@firstoftwo
 \else \expandafter \@secondoftwo
 \fi
}%
\providecommand \@ifx [1]{%
 \ifx #1\expandafter \@firstoftwo
 \else \expandafter \@secondoftwo
 \fi
}%
\providecommand \natexlab [1]{#1}%
\providecommand \enquote  [1]{``#1''}%
\providecommand \bibnamefont  [1]{#1}%
\providecommand \bibfnamefont [1]{#1}%
\providecommand \citenamefont [1]{#1}%
\providecommand \href@noop [0]{\@secondoftwo}%
\providecommand \href [0]{\begingroup \@sanitize@url \@href}%
\providecommand \@href[1]{\@@startlink{#1}\@@href}%
\providecommand \@@href[1]{\endgroup#1\@@endlink}%
\providecommand \@sanitize@url [0]{\catcode `\\12\catcode `\$12\catcode
  `\&12\catcode `\#12\catcode `\^12\catcode `\_12\catcode `\%12\relax}%
\providecommand \@@startlink[1]{}%
\providecommand \@@endlink[0]{}%
\providecommand \url  [0]{\begingroup\@sanitize@url \@url }%
\providecommand \@url [1]{\endgroup\@href {#1}{\urlprefix }}%
\providecommand \urlprefix  [0]{URL }%
\providecommand \Eprint [0]{\href }%
\providecommand \doibase [0]{http://dx.doi.org/}%
\providecommand \selectlanguage [0]{\@gobble}%
\providecommand \bibinfo  [0]{\@secondoftwo}%
\providecommand \bibfield  [0]{\@secondoftwo}%
\providecommand \translation [1]{[#1]}%
\providecommand \BibitemOpen [0]{}%
\providecommand \bibitemStop [0]{}%
\providecommand \bibitemNoStop [0]{.\EOS\space}%
\providecommand \EOS [0]{\spacefactor3000\relax}%
\providecommand \BibitemShut  [1]{\csname bibitem#1\endcsname}%
\let\auto@bib@innerbib\@empty
%</preamble>
\bibitem [{\citenamefont {Hasan}\ and\ \citenamefont
  {Kane}(2010)}]{TI_review_Hasan10}%
  \BibitemOpen
  \bibfield  {author} {\bibinfo {author} {\bibfnamefont {M.~Z.}\ \bibnamefont
  {Hasan}}\ and\ \bibinfo {author} {\bibfnamefont {C.~L.}\ \bibnamefont
  {Kane}},\ }\href {\doibase 10.1103/RevModPhys.82.3045} {\bibfield  {journal}
  {\bibinfo  {journal} {Rev. Mod. Phys.}\ }\textbf {\bibinfo {volume} {82}},\
  \bibinfo {pages} {3045} (\bibinfo {year} {2010})}\BibitemShut {NoStop}%
\bibitem [{\citenamefont {Qi}\ and\ \citenamefont
  {Zhang}(2011)}]{TI_review_Qi10}%
  \BibitemOpen
  \bibfield  {author} {\bibinfo {author} {\bibfnamefont {X.-L.}\ \bibnamefont
  {Qi}}\ and\ \bibinfo {author} {\bibfnamefont {S.-C.}\ \bibnamefont {Zhang}},\
  }\href {\doibase 10.1103/RevModPhys.83.1057} {\bibfield  {journal} {\bibinfo
  {journal} {Rev. Mod. Phys.}\ }\textbf {\bibinfo {volume} {83}},\ \bibinfo
  {pages} {1057} (\bibinfo {year} {2011})}\BibitemShut {NoStop}%
\bibitem [{\citenamefont {Schnyder}\ \emph {et~al.}(2008)\citenamefont
  {Schnyder}, \citenamefont {Ryu}, \citenamefont {Furusaki},\ and\
  \citenamefont {Ludwig}}]{Schnyder_classification_free_2008}%
  \BibitemOpen
  \bibfield  {author} {\bibinfo {author} {\bibfnamefont {A.~P.}\ \bibnamefont
  {Schnyder}}, \bibinfo {author} {\bibfnamefont {S.}~\bibnamefont {Ryu}},
  \bibinfo {author} {\bibfnamefont {A.}~\bibnamefont {Furusaki}}, \ and\
  \bibinfo {author} {\bibfnamefont {A.~W.~W.}\ \bibnamefont {Ludwig}},\ }\href
  {\doibase 10.1103/PhysRevB.78.195125} {\bibfield  {journal} {\bibinfo
  {journal} {Phys. Rev. B}\ }\textbf {\bibinfo {volume} {78}},\ \bibinfo
  {pages} {195125} (\bibinfo {year} {2008})}\BibitemShut {NoStop}%
\bibitem [{\citenamefont {Kitaev}(2009)}]{Kitaev_classification_free_2009}%
  \BibitemOpen
  \bibfield  {author} {\bibinfo {author} {\bibfnamefont {A.}~\bibnamefont
  {Kitaev}},\ }\href {\doibase 10.1063/1.3149495} {\bibfield  {journal}
  {\bibinfo  {journal} {AIP Conf. Proc.}\ }\textbf {\bibinfo {volume} {1134}},\
  \bibinfo {pages} {22} (\bibinfo {year} {2009})}\BibitemShut {NoStop}%
\bibitem [{\citenamefont {Ryu}\ \emph {et~al.}(2010)\citenamefont {Ryu},
  \citenamefont {Schnyder}, \citenamefont {Furusaki},\ and\ \citenamefont
  {Ludwig}}]{Ryu_classification_free_2010}%
  \BibitemOpen
  \bibfield  {author} {\bibinfo {author} {\bibfnamefont {S.}~\bibnamefont
  {Ryu}}, \bibinfo {author} {\bibfnamefont {A.~P.}\ \bibnamefont {Schnyder}},
  \bibinfo {author} {\bibfnamefont {A.}~\bibnamefont {Furusaki}}, \ and\
  \bibinfo {author} {\bibfnamefont {A.~W.~W.}\ \bibnamefont {Ludwig}},\ }\href
  {http://stacks.iop.org/1367-2630/12/i=6/a=065010} {\bibfield  {journal}
  {\bibinfo  {journal} {New J. Phys.}\ }\textbf {\bibinfo {volume} {12}},\
  \bibinfo {pages} {065010} (\bibinfo {year} {2010})}\BibitemShut {NoStop}%
\bibitem [{\citenamefont {Bernevig}\ \emph {et~al.}(2006)\citenamefont
  {Bernevig}, \citenamefont {Hughes},\ and\ \citenamefont
  {Zhang}}]{HgTe_Bernevig06}%
  \BibitemOpen
  \bibfield  {author} {\bibinfo {author} {\bibfnamefont {B.~A.}\ \bibnamefont
  {Bernevig}}, \bibinfo {author} {\bibfnamefont {T.~L.}\ \bibnamefont
  {Hughes}}, \ and\ \bibinfo {author} {\bibfnamefont {S.-C.}\ \bibnamefont
  {Zhang}},\ }\href {\doibase 10.1126/science.1133734} {\bibfield  {journal}
  {\bibinfo  {journal} {Science}\ }\textbf {\bibinfo {volume} {314}},\ \bibinfo
  {pages} {1757} (\bibinfo {year} {2006})},\ \Eprint
  {http://arxiv.org/abs/http://science.sciencemag.org/content/314/5806/1757.full.pdf}
  {http://science.sciencemag.org/content/314/5806/1757.full.pdf} \BibitemShut
  {NoStop}%
\bibitem [{\citenamefont {K{\"o}nig}\ \emph {et~al.}(2007)\citenamefont
  {K{\"o}nig}, \citenamefont {Wiedmann}, \citenamefont {Br{\"u}ne},
  \citenamefont {Roth}, \citenamefont {Buhmann}, \citenamefont {Molenkamp},
  \citenamefont {Qi},\ and\ \citenamefont {Zhang}}]{exp_2D-QW_MKonig_2007}%
  \BibitemOpen
  \bibfield  {author} {\bibinfo {author} {\bibfnamefont {M.}~\bibnamefont
  {K{\"o}nig}}, \bibinfo {author} {\bibfnamefont {S.}~\bibnamefont {Wiedmann}},
  \bibinfo {author} {\bibfnamefont {C.}~\bibnamefont {Br{\"u}ne}}, \bibinfo
  {author} {\bibfnamefont {A.}~\bibnamefont {Roth}}, \bibinfo {author}
  {\bibfnamefont {H.}~\bibnamefont {Buhmann}}, \bibinfo {author} {\bibfnamefont
  {L.~W.}\ \bibnamefont {Molenkamp}}, \bibinfo {author} {\bibfnamefont {X.-L.}\
  \bibnamefont {Qi}}, \ and\ \bibinfo {author} {\bibfnamefont {S.-C.}\
  \bibnamefont {Zhang}},\ }\href {\doibase 10.1126/science.1148047} {\bibfield
  {journal} {\bibinfo  {journal} {Science}\ }\textbf {\bibinfo {volume}
  {318}},\ \bibinfo {pages} {766} (\bibinfo {year} {2007})}\BibitemShut
  {NoStop}%
\bibitem [{\citenamefont {Hsieh}\ \emph {et~al.}(2008)\citenamefont {Hsieh},
  \citenamefont {Qian}, \citenamefont {Wray}, \citenamefont {Xia},
  \citenamefont {Hor}, \citenamefont {Cava},\ and\ \citenamefont
  {Hasan}}]{exp_3D-bismuth_YXia_2008}%
  \BibitemOpen
  \bibfield  {author} {\bibinfo {author} {\bibfnamefont {D.}~\bibnamefont
  {Hsieh}}, \bibinfo {author} {\bibfnamefont {D.}~\bibnamefont {Qian}},
  \bibinfo {author} {\bibfnamefont {L.}~\bibnamefont {Wray}}, \bibinfo {author}
  {\bibfnamefont {Y.}~\bibnamefont {Xia}}, \bibinfo {author} {\bibfnamefont
  {Y.~S.}\ \bibnamefont {Hor}}, \bibinfo {author} {\bibfnamefont
  {R.}~\bibnamefont {Cava}}, \ and\ \bibinfo {author} {\bibfnamefont {M.~Z.}\
  \bibnamefont {Hasan}},\ }\href@noop {} {\bibfield  {journal} {\bibinfo
  {journal} {Nature}\ }\textbf {\bibinfo {volume} {452}},\ \bibinfo {pages}
  {970} (\bibinfo {year} {2008})}\BibitemShut {NoStop}%
\bibitem [{\citenamefont {Chen}\ \emph {et~al.}(2009)\citenamefont {Chen},
  \citenamefont {Analytis}, \citenamefont {Chu}, \citenamefont {Liu},
  \citenamefont {Mo}, \citenamefont {Qi}, \citenamefont {Zhang}, \citenamefont
  {Lu}, \citenamefont {Dai}, \citenamefont {Fang}, \citenamefont {Zhang},
  \citenamefont {Fisher}, \citenamefont {Hussain},\ and\ \citenamefont
  {Shen}}]{exp_3D-bismuth_YLChen}%
  \BibitemOpen
  \bibfield  {author} {\bibinfo {author} {\bibfnamefont {Y.~L.}\ \bibnamefont
  {Chen}}, \bibinfo {author} {\bibfnamefont {J.~G.}\ \bibnamefont {Analytis}},
  \bibinfo {author} {\bibfnamefont {J.-H.}\ \bibnamefont {Chu}}, \bibinfo
  {author} {\bibfnamefont {Z.~K.}\ \bibnamefont {Liu}}, \bibinfo {author}
  {\bibfnamefont {S.-K.}\ \bibnamefont {Mo}}, \bibinfo {author} {\bibfnamefont
  {X.~L.}\ \bibnamefont {Qi}}, \bibinfo {author} {\bibfnamefont {H.~J.}\
  \bibnamefont {Zhang}}, \bibinfo {author} {\bibfnamefont {D.~H.}\ \bibnamefont
  {Lu}}, \bibinfo {author} {\bibfnamefont {X.}~\bibnamefont {Dai}}, \bibinfo
  {author} {\bibfnamefont {Z.}~\bibnamefont {Fang}}, \bibinfo {author}
  {\bibfnamefont {S.~C.}\ \bibnamefont {Zhang}}, \bibinfo {author}
  {\bibfnamefont {I.~R.}\ \bibnamefont {Fisher}}, \bibinfo {author}
  {\bibfnamefont {Z.}~\bibnamefont {Hussain}}, \ and\ \bibinfo {author}
  {\bibfnamefont {Z.-X.}\ \bibnamefont {Shen}},\ }\href {\doibase
  10.1126/science.1173034} {\bibfield  {journal} {\bibinfo  {journal}
  {Science}\ }\textbf {\bibinfo {volume} {325}},\ \bibinfo {pages} {178}
  (\bibinfo {year} {2009})}\BibitemShut {NoStop}%
\bibitem [{\citenamefont {Yi}\ \emph {et~al.}(2010)\citenamefont {Yi},
  \citenamefont {Ke}, \citenamefont {Cui-Zu}, \citenamefont {Can-Li},
  \citenamefont {Li-LiWang}, \citenamefont {Xi}, \citenamefont {Jin-Feng},
  \citenamefont {Zhong}, \citenamefont {Xi}, \citenamefont {Wen-Yu},
  \citenamefont {Shun-Qing}, \citenamefont {Qian}, \citenamefont {Xiao-Liang},
  \citenamefont {Shou-Cheng}, \citenamefont {Xu-Cun},\ and\ \citenamefont
  {Qi-Kun}}]{exp_3D_bismuth_Se_Zhang}%
  \BibitemOpen
  \bibfield  {author} {\bibinfo {author} {\bibfnamefont {Z.}~\bibnamefont
  {Yi}}, \bibinfo {author} {\bibfnamefont {H.}~\bibnamefont {Ke}}, \bibinfo
  {author} {\bibfnamefont {C.}~\bibnamefont {Cui-Zu}}, \bibinfo {author}
  {\bibfnamefont {S.}~\bibnamefont {Can-Li}}, \bibinfo {author} {\bibnamefont
  {Li-LiWang}}, \bibinfo {author} {\bibfnamefont {C.}~\bibnamefont {Xi}},
  \bibinfo {author} {\bibfnamefont {J.}~\bibnamefont {Jin-Feng}}, \bibinfo
  {author} {\bibfnamefont {F.}~\bibnamefont {Zhong}}, \bibinfo {author}
  {\bibfnamefont {D.}~\bibnamefont {Xi}}, \bibinfo {author} {\bibfnamefont
  {S.}~\bibnamefont {Wen-Yu}}, \bibinfo {author} {\bibfnamefont
  {S.}~\bibnamefont {Shun-Qing}}, \bibinfo {author} {\bibfnamefont
  {N.}~\bibnamefont {Qian}}, \bibinfo {author} {\bibfnamefont {Q.}~\bibnamefont
  {Xiao-Liang}}, \bibinfo {author} {\bibfnamefont {Z.}~\bibnamefont
  {Shou-Cheng}}, \bibinfo {author} {\bibfnamefont {M.}~\bibnamefont {Xu-Cun}},
  \ and\ \bibinfo {author} {\bibfnamefont {X.}~\bibnamefont {Qi-Kun}},\ }\href
  {http://search.ebscohost.com/login.aspx?direct=true&db=a9h&AN=52703016&lang=ja&site=ehost-live}
  {\bibfield  {journal} {\bibinfo  {journal} {Nature Physics}\ }\textbf
  {\bibinfo {volume} {6}},\ \bibinfo {pages} {584 } (\bibinfo {year}
  {2010})}\BibitemShut {NoStop}%
\bibitem [{\citenamefont {Sasaki}\ \emph {et~al.}(2011)\citenamefont {Sasaki},
  \citenamefont {Kriener}, \citenamefont {Segawa}, \citenamefont {Yada},
  \citenamefont {Tanaka}, \citenamefont {Sato},\ and\ \citenamefont
  {Ando}}]{exp-3D_TSC_Sasaki11}%
  \BibitemOpen
  \bibfield  {author} {\bibinfo {author} {\bibfnamefont {S.}~\bibnamefont
  {Sasaki}}, \bibinfo {author} {\bibfnamefont {M.}~\bibnamefont {Kriener}},
  \bibinfo {author} {\bibfnamefont {K.}~\bibnamefont {Segawa}}, \bibinfo
  {author} {\bibfnamefont {K.}~\bibnamefont {Yada}}, \bibinfo {author}
  {\bibfnamefont {Y.}~\bibnamefont {Tanaka}}, \bibinfo {author} {\bibfnamefont
  {M.}~\bibnamefont {Sato}}, \ and\ \bibinfo {author} {\bibfnamefont
  {Y.}~\bibnamefont {Ando}},\ }\href {\doibase 10.1103/PhysRevLett.107.217001}
  {\bibfield  {journal} {\bibinfo  {journal} {Phys. Rev. Lett.}\ }\textbf
  {\bibinfo {volume} {107}},\ \bibinfo {pages} {217001} (\bibinfo {year}
  {2011})}\BibitemShut {NoStop}%
\bibitem [{\citenamefont {Shitade}\ \emph {et~al.}(2009)\citenamefont
  {Shitade}, \citenamefont {Katsura}, \citenamefont
  {Kune\ifmmode~\check{s}\else \v{s}\fi{}}, \citenamefont {Qi}, \citenamefont
  {Zhang},\ and\ \citenamefont {Nagaosa}}]{NaIrO_Nagaosa09}%
  \BibitemOpen
  \bibfield  {author} {\bibinfo {author} {\bibfnamefont {A.}~\bibnamefont
  {Shitade}}, \bibinfo {author} {\bibfnamefont {H.}~\bibnamefont {Katsura}},
  \bibinfo {author} {\bibfnamefont {J.}~\bibnamefont
  {Kune\ifmmode~\check{s}\else \v{s}\fi{}}}, \bibinfo {author} {\bibfnamefont
  {X.-L.}\ \bibnamefont {Qi}}, \bibinfo {author} {\bibfnamefont {S.-C.}\
  \bibnamefont {Zhang}}, \ and\ \bibinfo {author} {\bibfnamefont
  {N.}~\bibnamefont {Nagaosa}},\ }\href {\doibase
  10.1103/PhysRevLett.102.256403} {\bibfield  {journal} {\bibinfo  {journal}
  {Phys. Rev. Lett.}\ }\textbf {\bibinfo {volume} {102}},\ \bibinfo {pages}
  {256403} (\bibinfo {year} {2009})}\BibitemShut {NoStop}%
\bibitem [{\citenamefont {Chadov}\ \emph {et~al.}(2010)\citenamefont {Chadov},
  \citenamefont {Qi}, \citenamefont {K{\"u}bler}, \citenamefont {Fecher},
  \citenamefont {Felser},\ and\ \citenamefont {Zhang}}]{Heusler_Chadov10}%
  \BibitemOpen
  \bibfield  {author} {\bibinfo {author} {\bibfnamefont {S.}~\bibnamefont
  {Chadov}}, \bibinfo {author} {\bibfnamefont {X.}~\bibnamefont {Qi}}, \bibinfo
  {author} {\bibfnamefont {J.}~\bibnamefont {K{\"u}bler}}, \bibinfo {author}
  {\bibfnamefont {G.~H.}\ \bibnamefont {Fecher}}, \bibinfo {author}
  {\bibfnamefont {C.}~\bibnamefont {Felser}}, \ and\ \bibinfo {author}
  {\bibfnamefont {S.~C.}\ \bibnamefont {Zhang}},\ }\href@noop {} {\bibfield
  {journal} {\bibinfo  {journal} {Nat. Mater.}\ }\textbf {\bibinfo {volume}
  {9}},\ \bibinfo {pages} {541} (\bibinfo {year} {2010})}\BibitemShut {NoStop}%
\bibitem [{\citenamefont {Lin}\ \emph {et~al.}(2010)\citenamefont {Lin},
  \citenamefont {Wray}, \citenamefont {Xia}, \citenamefont {Xu}, \citenamefont
  {Jia}, \citenamefont {Cava}, \citenamefont {Bansil},\ and\ \citenamefont
  {Hasan}}]{Heusler_Lin10}%
  \BibitemOpen
  \bibfield  {author} {\bibinfo {author} {\bibfnamefont {H.}~\bibnamefont
  {Lin}}, \bibinfo {author} {\bibfnamefont {L.~A.}\ \bibnamefont {Wray}},
  \bibinfo {author} {\bibfnamefont {Y.}~\bibnamefont {Xia}}, \bibinfo {author}
  {\bibfnamefont {S.}~\bibnamefont {Xu}}, \bibinfo {author} {\bibfnamefont
  {S.}~\bibnamefont {Jia}}, \bibinfo {author} {\bibfnamefont {R.~J.}\
  \bibnamefont {Cava}}, \bibinfo {author} {\bibfnamefont {A.}~\bibnamefont
  {Bansil}}, \ and\ \bibinfo {author} {\bibfnamefont {M.~Z.}\ \bibnamefont
  {Hasan}},\ }\href {\doibase 10.1038/nmat2771} {\bibfield  {journal} {\bibinfo
   {journal} {Nat. Mater.}\ }\textbf {\bibinfo {volume} {9}},\ \bibinfo {pages}
  {546} (\bibinfo {year} {2010})}\BibitemShut {NoStop}%
\bibitem [{\citenamefont {Yan}\ \emph {et~al.}(2012)\citenamefont {Yan},
  \citenamefont {M\"uchler}, \citenamefont {Qi}, \citenamefont {Zhang},\ and\
  \citenamefont {Felser}}]{skutterudites_Yan12}%
  \BibitemOpen
  \bibfield  {author} {\bibinfo {author} {\bibfnamefont {B.}~\bibnamefont
  {Yan}}, \bibinfo {author} {\bibfnamefont {L.}~\bibnamefont {M\"uchler}},
  \bibinfo {author} {\bibfnamefont {X.-L.}\ \bibnamefont {Qi}}, \bibinfo
  {author} {\bibfnamefont {S.-C.}\ \bibnamefont {Zhang}}, \ and\ \bibinfo
  {author} {\bibfnamefont {C.}~\bibnamefont {Felser}},\ }\href {\doibase
  10.1103/PhysRevB.85.165125} {\bibfield  {journal} {\bibinfo  {journal} {Phys.
  Rev. B}\ }\textbf {\bibinfo {volume} {85}},\ \bibinfo {pages} {165125}
  (\bibinfo {year} {2012})}\BibitemShut {NoStop}%
\bibitem [{\citenamefont {Takimoto}(2011)}]{Takimoto_2011}%
  \BibitemOpen
  \bibfield  {author} {\bibinfo {author} {\bibfnamefont {T.}~\bibnamefont
  {Takimoto}},\ }\href {\doibase 10.1143/JPSJ.80.123710} {\bibfield  {journal}
  {\bibinfo  {journal} {J. Phys. Soc. Jpn.}\ }\textbf {\bibinfo {volume}
  {80}},\ \bibinfo {pages} {123710} (\bibinfo {year} {2011})}\BibitemShut
  {NoStop}%
\bibitem [{\citenamefont {Lu}\ \emph {et~al.}(2013)\citenamefont {Lu},
  \citenamefont {Zhao}, \citenamefont {Weng}, \citenamefont {Fang},\ and\
  \citenamefont {Dai}}]{SmB6_Lu2013}%
  \BibitemOpen
  \bibfield  {author} {\bibinfo {author} {\bibfnamefont {F.}~\bibnamefont
  {Lu}}, \bibinfo {author} {\bibfnamefont {J.}~\bibnamefont {Zhao}}, \bibinfo
  {author} {\bibfnamefont {H.}~\bibnamefont {Weng}}, \bibinfo {author}
  {\bibfnamefont {Z.}~\bibnamefont {Fang}}, \ and\ \bibinfo {author}
  {\bibfnamefont {X.}~\bibnamefont {Dai}},\ }\href {\doibase
  10.1103/PhysRevLett.110.096401} {\bibfield  {journal} {\bibinfo  {journal}
  {Phys. Rev. Lett.}\ }\textbf {\bibinfo {volume} {110}},\ \bibinfo {pages}
  {096401} (\bibinfo {year} {2013})}\BibitemShut {NoStop}%
\bibitem [{\citenamefont {Kargarian}\ and\ \citenamefont
  {Fiete}(2013)}]{Kargarian_Fiete_TCIinOxide2013}%
  \BibitemOpen
  \bibfield  {author} {\bibinfo {author} {\bibfnamefont {M.}~\bibnamefont
  {Kargarian}}\ and\ \bibinfo {author} {\bibfnamefont {G.~A.}\ \bibnamefont
  {Fiete}},\ }\href {\doibase 10.1103/PhysRevLett.110.156403} {\bibfield
  {journal} {\bibinfo  {journal} {Phys. Rev. Lett.}\ }\textbf {\bibinfo
  {volume} {110}},\ \bibinfo {pages} {156403} (\bibinfo {year}
  {2013})}\BibitemShut {NoStop}%
\bibitem [{\citenamefont {Hsieh}\ \emph
  {et~al.}(2014{\natexlab{a}})\citenamefont {Hsieh}, \citenamefont {Liu},\ and\
  \citenamefont {Fu}}]{Hsieh_TCIinOxides2014}%
  \BibitemOpen
  \bibfield  {author} {\bibinfo {author} {\bibfnamefont {T.~H.}\ \bibnamefont
  {Hsieh}}, \bibinfo {author} {\bibfnamefont {J.}~\bibnamefont {Liu}}, \ and\
  \bibinfo {author} {\bibfnamefont {L.}~\bibnamefont {Fu}},\ }\href {\doibase
  10.1103/PhysRevB.90.081112} {\bibfield  {journal} {\bibinfo  {journal} {Phys.
  Rev. B}\ }\textbf {\bibinfo {volume} {90}},\ \bibinfo {pages} {081112}
  (\bibinfo {year} {2014}{\natexlab{a}})}\BibitemShut {NoStop}%
\bibitem [{\citenamefont {Weng}\ \emph {et~al.}(2014)\citenamefont {Weng},
  \citenamefont {Zhao}, \citenamefont {Wang}, \citenamefont {Fang},\ and\
  \citenamefont {Dai}}]{Weng_Dai_TCIinYbB12_2014}%
  \BibitemOpen
  \bibfield  {author} {\bibinfo {author} {\bibfnamefont {H.}~\bibnamefont
  {Weng}}, \bibinfo {author} {\bibfnamefont {J.}~\bibnamefont {Zhao}}, \bibinfo
  {author} {\bibfnamefont {Z.}~\bibnamefont {Wang}}, \bibinfo {author}
  {\bibfnamefont {Z.}~\bibnamefont {Fang}}, \ and\ \bibinfo {author}
  {\bibfnamefont {X.}~\bibnamefont {Dai}},\ }\href {\doibase
  10.1103/PhysRevLett.112.016403} {\bibfield  {journal} {\bibinfo  {journal}
  {Phys. Rev. Lett.}\ }\textbf {\bibinfo {volume} {112}},\ \bibinfo {pages}
  {016403} (\bibinfo {year} {2014})}\BibitemShut {NoStop}%
\bibitem [{\citenamefont {Fidkowski}\ and\ \citenamefont
  {Kitaev}(2010)}]{Z_to_Zn_Fidkowski_10}%
  \BibitemOpen
  \bibfield  {author} {\bibinfo {author} {\bibfnamefont {L.}~\bibnamefont
  {Fidkowski}}\ and\ \bibinfo {author} {\bibfnamefont {A.}~\bibnamefont
  {Kitaev}},\ }\href {\doibase 10.1103/PhysRevB.81.134509} {\bibfield
  {journal} {\bibinfo  {journal} {Phys. Rev. B}\ }\textbf {\bibinfo {volume}
  {81}},\ \bibinfo {pages} {134509} (\bibinfo {year} {2010})}\BibitemShut
  {NoStop}%
\bibitem [{\citenamefont {Fidkowski}\ and\ \citenamefont
  {Kitaev}(2011)}]{Fidkowski_1Dclassificatin_11}%
  \BibitemOpen
  \bibfield  {author} {\bibinfo {author} {\bibfnamefont {L.}~\bibnamefont
  {Fidkowski}}\ and\ \bibinfo {author} {\bibfnamefont {A.}~\bibnamefont
  {Kitaev}},\ }\href {\doibase 10.1103/PhysRevB.83.075103} {\bibfield
  {journal} {\bibinfo  {journal} {Phys. Rev. B}\ }\textbf {\bibinfo {volume}
  {83}},\ \bibinfo {pages} {075103} (\bibinfo {year} {2011})}\BibitemShut
  {NoStop}%
\bibitem [{\citenamefont {Turner}\ \emph {et~al.}(2011)\citenamefont {Turner},
  \citenamefont {Pollmann},\ and\ \citenamefont {Berg}}]{Turner11}%
  \BibitemOpen
  \bibfield  {author} {\bibinfo {author} {\bibfnamefont {A.~M.}\ \bibnamefont
  {Turner}}, \bibinfo {author} {\bibfnamefont {F.}~\bibnamefont {Pollmann}}, \
  and\ \bibinfo {author} {\bibfnamefont {E.}~\bibnamefont {Berg}},\ }\href
  {\doibase 10.1103/PhysRevB.83.075102} {\bibfield  {journal} {\bibinfo
  {journal} {Phys. Rev. B}\ }\textbf {\bibinfo {volume} {83}},\ \bibinfo
  {pages} {075102} (\bibinfo {year} {2011})}\BibitemShut {NoStop}%
\bibitem [{\citenamefont {Lu}\ and\ \citenamefont
  {Vishwanath}(2012)}]{Lu_CS_2011}%
  \BibitemOpen
  \bibfield  {author} {\bibinfo {author} {\bibfnamefont {Y.-M.}\ \bibnamefont
  {Lu}}\ and\ \bibinfo {author} {\bibfnamefont {A.}~\bibnamefont
  {Vishwanath}},\ }\href {\doibase 10.1103/PhysRevB.86.125119} {\bibfield
  {journal} {\bibinfo  {journal} {Phys. Rev. B}\ }\textbf {\bibinfo {volume}
  {86}},\ \bibinfo {pages} {125119} (\bibinfo {year} {2012})}\BibitemShut
  {NoStop}%
\bibitem [{\citenamefont {Levin}\ and\ \citenamefont
  {Stern}(2012)}]{Levin_CS_2012}%
  \BibitemOpen
  \bibfield  {author} {\bibinfo {author} {\bibfnamefont {M.}~\bibnamefont
  {Levin}}\ and\ \bibinfo {author} {\bibfnamefont {A.}~\bibnamefont {Stern}},\
  }\href {\doibase 10.1103/PhysRevB.86.115131} {\bibfield  {journal} {\bibinfo
  {journal} {Phys. Rev. B}\ }\textbf {\bibinfo {volume} {86}},\ \bibinfo
  {pages} {115131} (\bibinfo {year} {2012})}\BibitemShut {NoStop}%
\bibitem [{\citenamefont {Yao}\ and\ \citenamefont
  {Ryu}(2013)}]{YaoRyu_Z_to_Z8_2013}%
  \BibitemOpen
  \bibfield  {author} {\bibinfo {author} {\bibfnamefont {H.}~\bibnamefont
  {Yao}}\ and\ \bibinfo {author} {\bibfnamefont {S.}~\bibnamefont {Ryu}},\
  }\href {\doibase 10.1103/PhysRevB.88.064507} {\bibfield  {journal} {\bibinfo
  {journal} {Phys. Rev. B}\ }\textbf {\bibinfo {volume} {88}},\ \bibinfo
  {pages} {064507} (\bibinfo {year} {2013})}\BibitemShut {NoStop}%
\bibitem [{\citenamefont {Ryu}\ and\ \citenamefont
  {Zhang}(2012)}]{Ryu_Z_to_Z8_2013}%
  \BibitemOpen
  \bibfield  {author} {\bibinfo {author} {\bibfnamefont {S.}~\bibnamefont
  {Ryu}}\ and\ \bibinfo {author} {\bibfnamefont {S.-C.}\ \bibnamefont
  {Zhang}},\ }\href {\doibase 10.1103/PhysRevB.85.245132} {\bibfield  {journal}
  {\bibinfo  {journal} {Phys. Rev. B}\ }\textbf {\bibinfo {volume} {85}},\
  \bibinfo {pages} {245132} (\bibinfo {year} {2012})}\BibitemShut {NoStop}%
\bibitem [{\citenamefont {Hsieh}\ \emph
  {et~al.}(2014{\natexlab{b}})\citenamefont {Hsieh}, \citenamefont {Morimoto},\
  and\ \citenamefont {Ryu}}]{Hsieh_CS_CPT_2014}%
  \BibitemOpen
  \bibfield  {author} {\bibinfo {author} {\bibfnamefont {C.-T.}\ \bibnamefont
  {Hsieh}}, \bibinfo {author} {\bibfnamefont {T.}~\bibnamefont {Morimoto}}, \
  and\ \bibinfo {author} {\bibfnamefont {S.}~\bibnamefont {Ryu}},\ }\href
  {\doibase 10.1103/PhysRevB.90.245111} {\bibfield  {journal} {\bibinfo
  {journal} {Phys. Rev. B}\ }\textbf {\bibinfo {volume} {90}},\ \bibinfo
  {pages} {245111} (\bibinfo {year} {2014}{\natexlab{b}})}\BibitemShut
  {NoStop}%
\bibitem [{\citenamefont {Wang}\ \emph {et~al.}(2014)\citenamefont {Wang},
  \citenamefont {Potter},\ and\ \citenamefont
  {Senthil}}]{Wang_Potter_Senthil2014}%
  \BibitemOpen
  \bibfield  {author} {\bibinfo {author} {\bibfnamefont {C.}~\bibnamefont
  {Wang}}, \bibinfo {author} {\bibfnamefont {A.~C.}\ \bibnamefont {Potter}}, \
  and\ \bibinfo {author} {\bibfnamefont {T.}~\bibnamefont {Senthil}},\ }\href
  {\doibase 10.1126/science.1243326} {\bibfield  {journal} {\bibinfo  {journal}
  {Science}\ }\textbf {\bibinfo {volume} {343}},\ \bibinfo {pages} {629}
  (\bibinfo {year} {2014})}\BibitemShut {NoStop}%
\bibitem [{\citenamefont {Wang}\ and\ \citenamefont
  {Senthil}(2014)}]{Wang_Senthil2014}%
  \BibitemOpen
  \bibfield  {author} {\bibinfo {author} {\bibfnamefont {C.}~\bibnamefont
  {Wang}}\ and\ \bibinfo {author} {\bibfnamefont {T.}~\bibnamefont {Senthil}},\
  }\href {\doibase 10.1103/PhysRevB.89.195124} {\bibfield  {journal} {\bibinfo
  {journal} {Phys. Rev. B}\ }\textbf {\bibinfo {volume} {89}},\ \bibinfo
  {pages} {195124} (\bibinfo {year} {2014})}\BibitemShut {NoStop}%
\bibitem [{\citenamefont {You}\ and\ \citenamefont {Xu}(2014)}]{You_Cenke2014}%
  \BibitemOpen
  \bibfield  {author} {\bibinfo {author} {\bibfnamefont {Y.-Z.}\ \bibnamefont
  {You}}\ and\ \bibinfo {author} {\bibfnamefont {C.}~\bibnamefont {Xu}},\
  }\href {\doibase 10.1103/PhysRevB.90.245120} {\bibfield  {journal} {\bibinfo
  {journal} {Phys. Rev. B}\ }\textbf {\bibinfo {volume} {90}},\ \bibinfo
  {pages} {245120} (\bibinfo {year} {2014})}\BibitemShut {NoStop}%
\bibitem [{\citenamefont {Isobe}\ and\ \citenamefont
  {Fu}(2015)}]{Isobe_Fu2015}%
  \BibitemOpen
  \bibfield  {author} {\bibinfo {author} {\bibfnamefont {H.}~\bibnamefont
  {Isobe}}\ and\ \bibinfo {author} {\bibfnamefont {L.}~\bibnamefont {Fu}},\
  }\href@noop {} {\bibfield  {journal} {\bibinfo  {journal} {arXiv preprint
  arXiv:1502.06962}\ } (\bibinfo {year} {2015})}\BibitemShut {NoStop}%
\bibitem [{\citenamefont {Yoshida}\ and\ \citenamefont
  {Furusaki}(2015)}]{Yoshida2015}%
  \BibitemOpen
  \bibfield  {author} {\bibinfo {author} {\bibfnamefont {T.}~\bibnamefont
  {Yoshida}}\ and\ \bibinfo {author} {\bibfnamefont {A.}~\bibnamefont
  {Furusaki}},\ }\href {\doibase 10.1103/PhysRevB.92.085114} {\bibfield
  {journal} {\bibinfo  {journal} {Phys. Rev. B}\ }\textbf {\bibinfo {volume}
  {92}},\ \bibinfo {pages} {085114} (\bibinfo {year} {2015})}\BibitemShut
  {NoStop}%
\bibitem [{\citenamefont {Morimoto}\ \emph {et~al.}(2015)\citenamefont
  {Morimoto}, \citenamefont {Furusaki},\ and\ \citenamefont
  {Mudry}}]{Morimoto_2015}%
  \BibitemOpen
  \bibfield  {author} {\bibinfo {author} {\bibfnamefont {T.}~\bibnamefont
  {Morimoto}}, \bibinfo {author} {\bibfnamefont {A.}~\bibnamefont {Furusaki}},
  \ and\ \bibinfo {author} {\bibfnamefont {C.}~\bibnamefont {Mudry}},\ }\href
  {\doibase 10.1103/PhysRevB.92.125104} {\bibfield  {journal} {\bibinfo
  {journal} {Phys. Rev. B}\ }\textbf {\bibinfo {volume} {92}},\ \bibinfo
  {pages} {125104} (\bibinfo {year} {2015})}\BibitemShut {NoStop}%
\bibitem [{\citenamefont {Zhang}\ \emph {et~al.}(2016)\citenamefont {Zhang},
  \citenamefont {Xu},\ and\ \citenamefont {Liu}}]{TMKI_Zhang2016}%
  \BibitemOpen
  \bibfield  {author} {\bibinfo {author} {\bibfnamefont {R.-X.}\ \bibnamefont
  {Zhang}}, \bibinfo {author} {\bibfnamefont {C.}~\bibnamefont {Xu}}, \ and\
  \bibinfo {author} {\bibfnamefont {C.-X.}\ \bibnamefont {Liu}},\ }\href@noop
  {} {\bibfield  {journal} {\bibinfo  {journal} {arXiv preprint
  arXiv:1607.06073}\ } (\bibinfo {year} {2016})}\BibitemShut {NoStop}%
\bibitem [{Yos()}]{Yoshida_2017}%
  \BibitemOpen
  \href@noop {} {}\bibinfo {note} {{ T. Yoshida and N. Kawakami, to appear in
  PRB. }}\BibitemShut {NoStop}%
\bibitem [{\citenamefont {Mizukami}\ \emph {et~al.}(2011)\citenamefont
  {Mizukami}, \citenamefont {Shishido}, \citenamefont {Shibauchi},
  \citenamefont {Shimozawa}, \citenamefont {Yasumoto}, \citenamefont
  {Watanabe}, \citenamefont {Yamashita}, \citenamefont {Ikeda}, \citenamefont
  {Terashima}, \citenamefont {Kontani},\ and\ \citenamefont
  {Matsuda}}]{Mizukami_CeCoIn5YbCoIn5_11}%
  \BibitemOpen
  \bibfield  {author} {\bibinfo {author} {\bibfnamefont {Y.}~\bibnamefont
  {Mizukami}}, \bibinfo {author} {\bibfnamefont {H.}~\bibnamefont {Shishido}},
  \bibinfo {author} {\bibfnamefont {T.}~\bibnamefont {Shibauchi}}, \bibinfo
  {author} {\bibfnamefont {M.}~\bibnamefont {Shimozawa}}, \bibinfo {author}
  {\bibfnamefont {S.}~\bibnamefont {Yasumoto}}, \bibinfo {author}
  {\bibfnamefont {D.}~\bibnamefont {Watanabe}}, \bibinfo {author}
  {\bibfnamefont {M.}~\bibnamefont {Yamashita}}, \bibinfo {author}
  {\bibfnamefont {H.}~\bibnamefont {Ikeda}}, \bibinfo {author} {\bibfnamefont
  {T.}~\bibnamefont {Terashima}}, \bibinfo {author} {\bibfnamefont
  {H.}~\bibnamefont {Kontani}}, \ and\ \bibinfo {author} {\bibfnamefont
  {Y.}~\bibnamefont {Matsuda}},\ }\href
  {http://search.ebscohost.com/login.aspx?direct=true&db=a9h&AN=66954113&lang=ja&site=ehost-live}
  {\bibfield  {journal} {\bibinfo  {journal} {Nature Physics}\ }\textbf
  {\bibinfo {volume} {7}},\ \bibinfo {pages} {849 } (\bibinfo {year}
  {2011})}\BibitemShut {NoStop}%
\bibitem [{\citenamefont {Goh}\ \emph {et~al.}(2012)\citenamefont {Goh},
  \citenamefont {Mizukami}, \citenamefont {Shishido}, \citenamefont {Watanabe},
  \citenamefont {Yasumoto}, \citenamefont {Shimozawa}, \citenamefont
  {Yamashita}, \citenamefont {Terashima}, \citenamefont {Yanase}, \citenamefont
  {Shibauchi}, \citenamefont {Buzdin},\ and\ \citenamefont
  {Matsuda}}]{Goh_superlattice12}%
  \BibitemOpen
  \bibfield  {author} {\bibinfo {author} {\bibfnamefont {S.~K.}\ \bibnamefont
  {Goh}}, \bibinfo {author} {\bibfnamefont {Y.}~\bibnamefont {Mizukami}},
  \bibinfo {author} {\bibfnamefont {H.}~\bibnamefont {Shishido}}, \bibinfo
  {author} {\bibfnamefont {D.}~\bibnamefont {Watanabe}}, \bibinfo {author}
  {\bibfnamefont {S.}~\bibnamefont {Yasumoto}}, \bibinfo {author}
  {\bibfnamefont {M.}~\bibnamefont {Shimozawa}}, \bibinfo {author}
  {\bibfnamefont {M.}~\bibnamefont {Yamashita}}, \bibinfo {author}
  {\bibfnamefont {T.}~\bibnamefont {Terashima}}, \bibinfo {author}
  {\bibfnamefont {Y.}~\bibnamefont {Yanase}}, \bibinfo {author} {\bibfnamefont
  {T.}~\bibnamefont {Shibauchi}}, \bibinfo {author} {\bibfnamefont {A.~I.}\
  \bibnamefont {Buzdin}}, \ and\ \bibinfo {author} {\bibfnamefont
  {Y.}~\bibnamefont {Matsuda}},\ }\href {\doibase
  10.1103/PhysRevLett.109.157006} {\bibfield  {journal} {\bibinfo  {journal}
  {Phys. Rev. Lett.}\ }\textbf {\bibinfo {volume} {109}},\ \bibinfo {pages}
  {157006} (\bibinfo {year} {2012})}\BibitemShut {NoStop}%
\bibitem [{\citenamefont {Shimozawa}\ \emph {et~al.}(2014)\citenamefont
  {Shimozawa}, \citenamefont {Goh}, \citenamefont {Endo}, \citenamefont
  {Kobayashi}, \citenamefont {Watashige}, \citenamefont {Mizukami},
  \citenamefont {Ikeda}, \citenamefont {Shishido}, \citenamefont {Yanase},
  \citenamefont {Terashima}, \citenamefont {Shibauchi},\ and\ \citenamefont
  {Matsuda}}]{Shimozawa_superlattice_PRL14}%
  \BibitemOpen
  \bibfield  {author} {\bibinfo {author} {\bibfnamefont {M.}~\bibnamefont
  {Shimozawa}}, \bibinfo {author} {\bibfnamefont {S.~K.}\ \bibnamefont {Goh}},
  \bibinfo {author} {\bibfnamefont {R.}~\bibnamefont {Endo}}, \bibinfo {author}
  {\bibfnamefont {R.}~\bibnamefont {Kobayashi}}, \bibinfo {author}
  {\bibfnamefont {T.}~\bibnamefont {Watashige}}, \bibinfo {author}
  {\bibfnamefont {Y.}~\bibnamefont {Mizukami}}, \bibinfo {author}
  {\bibfnamefont {H.}~\bibnamefont {Ikeda}}, \bibinfo {author} {\bibfnamefont
  {H.}~\bibnamefont {Shishido}}, \bibinfo {author} {\bibfnamefont
  {Y.}~\bibnamefont {Yanase}}, \bibinfo {author} {\bibfnamefont
  {T.}~\bibnamefont {Terashima}}, \bibinfo {author} {\bibfnamefont
  {T.}~\bibnamefont {Shibauchi}}, \ and\ \bibinfo {author} {\bibfnamefont
  {Y.}~\bibnamefont {Matsuda}},\ }\href {\doibase
  10.1103/PhysRevLett.112.156404} {\bibfield  {journal} {\bibinfo  {journal}
  {Phys. Rev. Lett.}\ }\textbf {\bibinfo {volume} {112}},\ \bibinfo {pages}
  {156404} (\bibinfo {year} {2014})}\BibitemShut {NoStop}%
\bibitem [{\citenamefont {Shimozawa}\ \emph {et~al.}(2016)\citenamefont
  {Shimozawa}, \citenamefont {Goh}, \citenamefont {Shibauchi},\ and\
  \citenamefont {Matsuda}}]{Shimozawa_superlattice_RPP2016}%
  \BibitemOpen
  \bibfield  {author} {\bibinfo {author} {\bibfnamefont {M.}~\bibnamefont
  {Shimozawa}}, \bibinfo {author} {\bibfnamefont {S.~K.}\ \bibnamefont {Goh}},
  \bibinfo {author} {\bibfnamefont {T.}~\bibnamefont {Shibauchi}}, \ and\
  \bibinfo {author} {\bibfnamefont {Y.}~\bibnamefont {Matsuda}},\ }\href
  {http://stacks.iop.org/0034-4885/79/i=7/a=074503} {\bibfield  {journal}
  {\bibinfo  {journal} {Reports on Progress in Physics}\ }\textbf {\bibinfo
  {volume} {79}},\ \bibinfo {pages} {074503} (\bibinfo {year}
  {2016})}\BibitemShut {NoStop}%
\bibitem [{\citenamefont {Yoshida}\ \emph {et~al.}(2015)\citenamefont
  {Yoshida}, \citenamefont {Sigrist},\ and\ \citenamefont
  {Yanase}}]{superlattice_Yanase15}%
  \BibitemOpen
  \bibfield  {author} {\bibinfo {author} {\bibfnamefont {T.}~\bibnamefont
  {Yoshida}}, \bibinfo {author} {\bibfnamefont {M.}~\bibnamefont {Sigrist}}, \
  and\ \bibinfo {author} {\bibfnamefont {Y.}~\bibnamefont {Yanase}},\ }\href
  {\doibase 10.1103/PhysRevLett.115.027001} {\bibfield  {journal} {\bibinfo
  {journal} {Phys. Rev. Lett.}\ }\textbf {\bibinfo {volume} {115}},\ \bibinfo
  {pages} {027001} (\bibinfo {year} {2015})}\BibitemShut {NoStop}%
\bibitem [{\citenamefont {Yoshida}\ \emph {et~al.}()\citenamefont {Yoshida},
  \citenamefont {Sigrist},\ and\ \citenamefont
  {Yanase}}]{superlattice_Yanase16}%
  \BibitemOpen
  \bibfield  {author} {\bibinfo {author} {\bibfnamefont {T.}~\bibnamefont
  {Yoshida}}, \bibinfo {author} {\bibfnamefont {M.}~\bibnamefont {Sigrist}}, \
  and\ \bibinfo {author} {\bibfnamefont {Y.}~\bibnamefont {Yanase}},\
  }\href@noop {} {\bibinfo  {journal} {in preparation}\ }\BibitemShut {NoStop}%
\bibitem [{\citenamefont {She}\ and\ \citenamefont
  {Balatsky}(2012)}]{superlattice_proximity_12}%
  \BibitemOpen
\bibfield  {journal} {  }\bibfield  {author} {\bibinfo {author} {\bibfnamefont
  {J.-H.}\ \bibnamefont {She}}\ and\ \bibinfo {author} {\bibfnamefont {A.~V.}\
  \bibnamefont {Balatsky}},\ }\href {\doibase 10.1103/PhysRevLett.109.077002}
  {\bibfield  {journal} {\bibinfo  {journal} {Phys. Rev. Lett.}\ }\textbf
  {\bibinfo {volume} {109}},\ \bibinfo {pages} {077002} (\bibinfo {year}
  {2012})}\BibitemShut {NoStop}%
\bibitem [{\citenamefont {Yoshida}\ \emph {et~al.}(2012)\citenamefont
  {Yoshida}, \citenamefont {Sigrist},\ and\ \citenamefont
  {Yanase}}]{superlattice_Yanase_12}%
  \BibitemOpen
  \bibfield  {author} {\bibinfo {author} {\bibfnamefont {T.}~\bibnamefont
  {Yoshida}}, \bibinfo {author} {\bibfnamefont {M.}~\bibnamefont {Sigrist}}, \
  and\ \bibinfo {author} {\bibfnamefont {Y.}~\bibnamefont {Yanase}},\ }\href
  {\doibase 10.1103/PhysRevB.86.134514} {\bibfield  {journal} {\bibinfo
  {journal} {Phys. Rev. B}\ }\textbf {\bibinfo {volume} {86}},\ \bibinfo
  {pages} {134514} (\bibinfo {year} {2012})}\BibitemShut {NoStop}%
\bibitem [{\citenamefont {Tayama}\ \emph {et~al.}(2002)\citenamefont {Tayama},
  \citenamefont {Harita}, \citenamefont {Sakakibara}, \citenamefont {Haga},
  \citenamefont {Shishido}, \citenamefont {Settai},\ and\ \citenamefont
  {Onuki}}]{Onuki_CeCoIn5_Hc2_02}%
  \BibitemOpen
  \bibfield  {author} {\bibinfo {author} {\bibfnamefont {T.}~\bibnamefont
  {Tayama}}, \bibinfo {author} {\bibfnamefont {A.}~\bibnamefont {Harita}},
  \bibinfo {author} {\bibfnamefont {T.}~\bibnamefont {Sakakibara}}, \bibinfo
  {author} {\bibfnamefont {Y.}~\bibnamefont {Haga}}, \bibinfo {author}
  {\bibfnamefont {H.}~\bibnamefont {Shishido}}, \bibinfo {author}
  {\bibfnamefont {R.}~\bibnamefont {Settai}}, \ and\ \bibinfo {author}
  {\bibfnamefont {Y.}~\bibnamefont {Onuki}},\ }\href {\doibase
  10.1103/PhysRevB.65.180504} {\bibfield  {journal} {\bibinfo  {journal} {Phys.
  Rev. B}\ }\textbf {\bibinfo {volume} {65}},\ \bibinfo {pages} {180504}
  (\bibinfo {year} {2002})}\BibitemShut {NoStop}%
\bibitem [{\citenamefont {Maruyama}\ \emph {et~al.}(2012)\citenamefont
  {Maruyama}, \citenamefont {Sigrist},\ and\ \citenamefont
  {Yanase}}]{Maruyama_superlattice_12}%
  \BibitemOpen
  \bibfield  {author} {\bibinfo {author} {\bibfnamefont {D.}~\bibnamefont
  {Maruyama}}, \bibinfo {author} {\bibfnamefont {M.}~\bibnamefont {Sigrist}}, \
  and\ \bibinfo {author} {\bibfnamefont {Y.}~\bibnamefont {Yanase}},\ }\href
  {\doibase 10.1143/JPSJ.81.034702} {\bibfield  {journal} {\bibinfo  {journal}
  {Journal of the Physical Society of Japan}\ }\textbf {\bibinfo {volume}
  {81}},\ \bibinfo {pages} {034702} (\bibinfo {year} {2012})},\ \Eprint
  {http://arxiv.org/abs/http://dx.doi.org/10.1143/JPSJ.81.034702}
  {http://dx.doi.org/10.1143/JPSJ.81.034702} \BibitemShut {NoStop}%
\bibitem [{\citenamefont {Matsuda}\ \emph {et~al.}(2006)\citenamefont
  {Matsuda}, \citenamefont {Izawa},\ and\ \citenamefont
  {Vekhter}}]{angle_resolved_thermal_Matsuda_06}%
  \BibitemOpen
  \bibfield  {author} {\bibinfo {author} {\bibfnamefont {Y.}~\bibnamefont
  {Matsuda}}, \bibinfo {author} {\bibfnamefont {K.}~\bibnamefont {Izawa}}, \
  and\ \bibinfo {author} {\bibfnamefont {I.}~\bibnamefont {Vekhter}},\ }\href
  {http://stacks.iop.org/0953-8984/18/i=44/a=R01} {\bibfield  {journal}
  {\bibinfo  {journal} {Journal of Physics: Condensed Matter}\ }\textbf
  {\bibinfo {volume} {18}},\ \bibinfo {pages} {R705} (\bibinfo {year}
  {2006})}\BibitemShut {NoStop}%
\bibitem [{\citenamefont {Sigrist}\ and\ \citenamefont
  {Ueda}(1991)}]{SigristUeda_91}%
  \BibitemOpen
  \bibfield  {author} {\bibinfo {author} {\bibfnamefont {M.}~\bibnamefont
  {Sigrist}}\ and\ \bibinfo {author} {\bibfnamefont {K.}~\bibnamefont {Ueda}},\
  }\href {\doibase 10.1103/RevModPhys.63.239} {\bibfield  {journal} {\bibinfo
  {journal} {Rev. Mod. Phys.}\ }\textbf {\bibinfo {volume} {63}},\ \bibinfo
  {pages} {239} (\bibinfo {year} {1991})}\BibitemShut {NoStop}%
\bibitem [{sup()}]{supp}%
  \BibitemOpen
  \href@noop {} {}\bibinfo {note} {{ Supplementary Material. }}\BibitemShut
  {NoStop}%
\bibitem [{\citenamefont {Shiozaki}\ and\ \citenamefont
  {Sato}(2014)}]{shiozaki_classification_2014}%
  \BibitemOpen
  \bibfield  {author} {\bibinfo {author} {\bibfnamefont {K.}~\bibnamefont
  {Shiozaki}}\ and\ \bibinfo {author} {\bibfnamefont {M.}~\bibnamefont
  {Sato}},\ }\href {\doibase 10.1103/PhysRevB.90.165114} {\bibfield  {journal}
  {\bibinfo  {journal} {Phys. Rev. B}\ }\textbf {\bibinfo {volume} {90}},\
  \bibinfo {pages} {165114} (\bibinfo {year} {2014})}\BibitemShut {NoStop}%
\bibitem [{\citenamefont {Fukui}\ \emph {et~al.}(2005)\citenamefont {Fukui},
  \citenamefont {Hatsugai},\ and\ \citenamefont {Suzuki}}]{Fukui_Hatsugai_05}%
  \BibitemOpen
  \bibfield  {author} {\bibinfo {author} {\bibfnamefont {T.}~\bibnamefont
  {Fukui}}, \bibinfo {author} {\bibfnamefont {Y.}~\bibnamefont {Hatsugai}}, \
  and\ \bibinfo {author} {\bibfnamefont {H.}~\bibnamefont {Suzuki}},\ }\href
  {\doibase 10.1143/JPSJ.74.1674} {\bibfield  {journal} {\bibinfo  {journal}
  {Journal of the Physical Society of Japan}\ }\textbf {\bibinfo {volume}
  {74}},\ \bibinfo {pages} {1674} (\bibinfo {year} {2005})},\ \Eprint
  {http://arxiv.org/abs/http://dx.doi.org/10.1143/JPSJ.74.1674}
  {http://dx.doi.org/10.1143/JPSJ.74.1674} \BibitemShut {NoStop}%
\bibitem [{\citenamefont {Haldane}(1995)}]{haldane_nullvector}%
  \BibitemOpen
  \bibfield  {author} {\bibinfo {author} {\bibfnamefont {F.}~\bibnamefont
  {Haldane}},\ }\href@noop {} {\bibfield  {journal} {\bibinfo  {journal} {Phys.
  Rev. Lett.}\ }\textbf {\bibinfo {volume} {74}},\ \bibinfo {pages} {2090}
  (\bibinfo {year} {1995})}\BibitemShut {NoStop}%
\bibitem [{\citenamefont {Mourik}\ \emph {et~al.}(2012)\citenamefont {Mourik},
  \citenamefont {Zuo}, \citenamefont {Frolov}, \citenamefont {Plissard},
  \citenamefont {Bakkers},\ and\ \citenamefont
  {Kouwenhoven}}]{Mourik_Majorana2012}%
  \BibitemOpen
  \bibfield  {author} {\bibinfo {author} {\bibfnamefont {V.}~\bibnamefont
  {Mourik}}, \bibinfo {author} {\bibfnamefont {K.}~\bibnamefont {Zuo}},
  \bibinfo {author} {\bibfnamefont {S.~M.}\ \bibnamefont {Frolov}}, \bibinfo
  {author} {\bibfnamefont {S.~R.}\ \bibnamefont {Plissard}}, \bibinfo {author}
  {\bibfnamefont {E.~P. A.~M.}\ \bibnamefont {Bakkers}}, \ and\ \bibinfo
  {author} {\bibfnamefont {L.~P.}\ \bibnamefont {Kouwenhoven}},\ }\href
  {\doibase 10.1126/science.1222360} {\bibfield  {journal} {\bibinfo  {journal}
  {Science}\ }\textbf {\bibinfo {volume} {336}},\ \bibinfo {pages} {1003}
  (\bibinfo {year} {2012})}\BibitemShut {NoStop}%
\bibitem [{\citenamefont {Nadj-Perge}\ \emph {et~al.}(2014)\citenamefont
  {Nadj-Perge}, \citenamefont {Drozdov}, \citenamefont {Li}, \citenamefont
  {Chen}, \citenamefont {Jeon}, \citenamefont {Seo}, \citenamefont {MacDonald},
  \citenamefont {Bernevig},\ and\ \citenamefont
  {Yazdani}}]{Nadj-Perge_Majorana2014}%
  \BibitemOpen
  \bibfield  {author} {\bibinfo {author} {\bibfnamefont {S.}~\bibnamefont
  {Nadj-Perge}}, \bibinfo {author} {\bibfnamefont {I.~K.}\ \bibnamefont
  {Drozdov}}, \bibinfo {author} {\bibfnamefont {J.}~\bibnamefont {Li}},
  \bibinfo {author} {\bibfnamefont {H.}~\bibnamefont {Chen}}, \bibinfo {author}
  {\bibfnamefont {S.}~\bibnamefont {Jeon}}, \bibinfo {author} {\bibfnamefont
  {J.}~\bibnamefont {Seo}}, \bibinfo {author} {\bibfnamefont {A.~H.}\
  \bibnamefont {MacDonald}}, \bibinfo {author} {\bibfnamefont {B.~A.}\
  \bibnamefont {Bernevig}}, \ and\ \bibinfo {author} {\bibfnamefont
  {A.}~\bibnamefont {Yazdani}},\ }\href {\doibase 10.1126/science.1259327}
  {\bibfield  {journal} {\bibinfo  {journal} {Science}\ }\textbf {\bibinfo
  {volume} {346}},\ \bibinfo {pages} {602} (\bibinfo {year} {2014})},\ \Eprint
  {http://arxiv.org/abs/http://science.sciencemag.org/content/346/6209/602.full.pdf}
  {http://science.sciencemag.org/content/346/6209/602.full.pdf} \BibitemShut
  {NoStop}%
\bibitem [{Mat()}]{Matsuda_discussion_2017}%
  \BibitemOpen
  \href@noop {} {}\bibinfo {note} {{ Y. Matsuda (private communication)
  }}\BibitemShut {NoStop}%
\bibitem [{\citenamefont {Allan}\ \emph {et~al.}(2013)\citenamefont {Allan},
  \citenamefont {Massee}, \citenamefont {Morr}, \citenamefont {Van~Dyke},
  \citenamefont {Rost}, \citenamefont {Mackenzie}, \citenamefont {Petrovic},\
  and\ \citenamefont {Davis}}]{STM_CeCoIn5_2013}%
  \BibitemOpen
  \bibfield  {author} {\bibinfo {author} {\bibfnamefont {M.}~\bibnamefont
  {Allan}}, \bibinfo {author} {\bibfnamefont {F.}~\bibnamefont {Massee}},
  \bibinfo {author} {\bibfnamefont {D.}~\bibnamefont {Morr}}, \bibinfo {author}
  {\bibfnamefont {J.}~\bibnamefont {Van~Dyke}}, \bibinfo {author}
  {\bibfnamefont {A.}~\bibnamefont {Rost}}, \bibinfo {author} {\bibfnamefont
  {A.}~\bibnamefont {Mackenzie}}, \bibinfo {author} {\bibfnamefont
  {C.}~\bibnamefont {Petrovic}}, \ and\ \bibinfo {author} {\bibfnamefont
  {J.}~\bibnamefont {Davis}},\ }\href@noop {} {\bibfield  {journal} {\bibinfo
  {journal} {Nature Physics}\ }\textbf {\bibinfo {volume} {9}},\ \bibinfo
  {pages} {468} (\bibinfo {year} {2013})}\BibitemShut {NoStop}%
\bibitem [{\citenamefont {Sidorov}\ \emph {et~al.}(2002)\citenamefont
  {Sidorov}, \citenamefont {Nicklas}, \citenamefont {Pagliuso}, \citenamefont
  {Sarrao}, \citenamefont {Bang}, \citenamefont {Balatsky},\ and\ \citenamefont
  {Thompson}}]{Sidorov_2002}%
  \BibitemOpen
  \bibfield  {author} {\bibinfo {author} {\bibfnamefont {V.~A.}\ \bibnamefont
  {Sidorov}}, \bibinfo {author} {\bibfnamefont {M.}~\bibnamefont {Nicklas}},
  \bibinfo {author} {\bibfnamefont {P.~G.}\ \bibnamefont {Pagliuso}}, \bibinfo
  {author} {\bibfnamefont {J.~L.}\ \bibnamefont {Sarrao}}, \bibinfo {author}
  {\bibfnamefont {Y.}~\bibnamefont {Bang}}, \bibinfo {author} {\bibfnamefont
  {A.~V.}\ \bibnamefont {Balatsky}}, \ and\ \bibinfo {author} {\bibfnamefont
  {J.~D.}\ \bibnamefont {Thompson}},\ }\href {\doibase
  10.1103/PhysRevLett.89.157004} {\bibfield  {journal} {\bibinfo  {journal}
  {Phys. Rev. Lett.}\ }\textbf {\bibinfo {volume} {89}},\ \bibinfo {pages}
  {157004} (\bibinfo {year} {2002})}\BibitemShut {NoStop}%
\end{thebibliography}
\end{document}